\documentclass[usenatbib]{mn2e}
\usepackage{graphicx}
\newcommand{\fracb}[2]{\left(\frac{#1}{#2}\right)}

\newcommand{\bfrac}[2]{\left(\frac{#1}{#2}\right)}

\newcommand{\text}[1]{\quad\mbox{#1}\quad}

\newcommand{\apj}{ApJ}

\newcommand{\apjl}{ApJ}
\newcommand{\mnras}{MNRAS}

\newcommand{\aap}{A\&A}

\newcommand{\sub}[1]{_{\mbox{\tiny #1}}}
\title{Close Binary Progenitors of Long Gamma Ray Bursts} 
\author[M.V.~Barkov and S.S. Komissarov]
{Maxim V.~Barkov,$^{1,2}$\thanks{E-Mail: bmv@maths.leeds.ac.uk} 
Serguei S.~Komissarov$^{1}$\thanks{E-Mail: serguei@maths.leeds.ac.uk} \\ 
$^{1}$Department of Applied Mathematics, The University of Leeds,
Leeds, LS2 9GT, UK\\
$^{2}$Space Research Institute, 84/32 Profsoyuznaya Street, Moscow
117997, Russia}

\begin{document}
\date{Received/Accepted}
\maketitle
                                                                              
%%%%%%%%%%%%%%%%%%%%%%%%%%%%%%%%%%%%%%%%%%%%%%%%%%
\begin{abstract}
%%%%%%%%%%%%%%%%%%%%%%%%%%%%%%%%%%%%%%%%%%%%%%%%%%
The strong dependence of the neutrino annihilation mechanism on the mass
accretion rate makes it difficult to explain the LGRBs with duration in excess
of 100 seconds as well as the precursors separated from the main gamma-ray
pulse by few hundreds of seconds.  Even more difficult is to explain the {\it
Swift} observations of the shallow decay phase and X-ray flares, if they
indeed indicate activity of the central engine for as long as $10^4$
seconds. These data suggest that some other, most likely magnetic mechanisms
have to be considered.  Since the efficiency of magnetic mechanisms does not
depend that much on the mass accretion rate, the magnetic models do not
require the development of accretion disk within the first few seconds of the
stellar collapse and hence do not require very rapidly rotating stellar cores
at the pre-supernova state.  This widens the range of potential LGRB
progenitors. In this paper, we re-examine the close binary scenario allowing
for the possibility of late development of accretion disks in the collapsar
model and investigate the available range of mass accretion rates, black hole
masses, and spins. We find that the black hole mass can be much higher than
$2-3M_\odot$, usually assumed in the collapsar model, and normally exceeds
half of the pre-supernova mass. The black hole spin is rather moderate,
$a=0.4-0.8$, but still high enough for the Blandford-Znajek mechanism to
remain efficient provided the magnetic field is sufficiently strong.  Our
numerical simulations confirm the possibility of magnetically driven stellar
explosions, in agreement with previous studies, but point towards the required
magnetic flux on the black hole horizon in excess of $10^{28}\mbox{G}\,\mbox{cm}^2$. 
At present, we cannot answer with certainty whether such a strong magnetic
field can be generated in the stellar interior.  Perhaps, the supernova
explosions associated with LGRBs are still neutrino-driven and their gamma-ray
signature is the precursors.  The supernova blast clears up escape channels
for the magnetically driven GRB jets, which may produce the main pulse. In
this scenario, the requirements on the magnetic field strength can be lowered.
A particularly interesting version of the binary progenitor involves merger of
a WR star with an ultra-compact companion, neutron star or black hole. In this
case we expect the formation of very long-lived accretion disks, that may
explain the phase of shallow decay and X-ray flares observed by {\it
Swift}. Similarly long-lived magnetic central engines are expected in the
current single star models of LGRB progenitors due to their assumed
exceptionally fast rotation.

\end{abstract}

\begin{keywords}
black hole physics -- accretion disks -- supernovae: general -- gamma-rays: bursts --
binaries: close  -- MHD -- relativity
\end{keywords}

%%%%%%%%%%%%%%%%%%%%%%%%%%%%%%%%%%%%%%%%%%%%%%%%%%
\section{Introduction}
\label{intro}

The nature of Gamma Ray Bursts (GRBs) remains one of the most intriguing
problems of modern astrophysics. It is now widely accepted that the gamma ray
emission is generated in ultrarelativistic jets but many basic questions
related both to the physics of these jets and to the mechanisms of their
production remain open. Although many promising theories have been developed
over the years since the discovery of GRBs, we are still some way out from
solid understanding of this phenomenon. For example, the observed connection
between the Long-duration Gamma Ray Bursts (LGRBs) and supernovae (SNe)
indicates that these bursts are connected to deaths of massive stars but the
details are not clear. In one model of LGRBs, the stellar collapse results in
a normal successful supernova explosion but the newly born neutron star (NS)
is very unusual.  It has both exceptionally high magnetic field, and for this
reason it is called a magnetar, and extremely rapid rotation
\citep[e.g.][]{U92,TCQ04,MTQ07,UM07}.  The powerful magnetohydrodynamic (MHD)
wind produced by such remnant is capable of both accelerating the supernova
shell above the expansion speed of normal supernovae, to the level of
hypernovae, and production of collimated ultra-relativistic polar jets
\citep{KB07,B09}.

In another model, the normal supernova explosion fails and the proto-neutron
star promptly collapses into a black hole (BH). However, the rapid rotation of
the stellar progenitor prevents the rest of the star from falling directly
into the black hole and a massive neutrino-cooled accretion disk is formed
instead.  This allows to turn the failed supernova into a successful stellar
explosion, as this disk can release enormous amounts of energy
\citep{W93,mw99}.  One way of ``utilising'' this energy is via the neutrino-
or magnetically-driven wind from the disk. Such wind is not expected to be
relativistic due to the high mass loading at its base. However, the polar
region just above the black hole is less likely to become mass-loaded by the
disk matter and can become relativistically hot via annihilation of neutrinos
and anti-neutrinos emitted by the disk. This opens a possibility of driving
ultra-relativistic LGRB jets in the collapsar scenario
\citep[e.g.][]{mw99,AIMGM00}.  However, the efficiency of this type of
neutrino heating is a very strong function of both the mass accretion rate and
the rotation rate of the central black hole \citep{pwf99,CB07,ZB09}.
According to the calculations by \citet{pwf99} for the black hole with the
rotation parameter $a=0.5$ the energy deposition rate via the neutrino
annihilation process drops from $L_{\nu\bar{\nu}}=4\times10^{48}$erg/s for
$\dot{M}=0.1 M_\odot$ to $L_{\nu\bar{\nu}}=6\times10^{44}$erg/s for
$\dot{M}=0.01 M_\odot$ and for the accretion rate of $\dot{M}=0.1 M_\odot$
from $L_{\nu\bar{\nu}}=2\times10^{51}$erg/s for $a=0.95$ to
$L_{\nu\bar{\nu}}=3\times10^{48}$erg/s for $a=0$.  Therefore, this version of
the collapsar model, similarly to the magnetar model, requires very rapid
rotation of the stellar core prior to the collapse so that the accretion disk
is formed early on, when the accretion rate is still high enough, and the BH
is born rapidly rotating. The results by \citet{bir07} suggest that
\citet{pwf99} may have overestimated the efficiency of neutrino mechanism for
high $a$. This is because the energy released by the disk powers not only the
outflow but also the flow into the BH.  As $a$ increases the inner boundary of
the disk moves closer to the BH and a larger fraction of the total
neutrino-antineutrino annihilation occurs in the region where the vector of
deposited momentum points towards the black hole.  In fact, \citet{bir07} find
that the efficiency of the neutrino annihilation mechanism peaks at
$a\simeq0.6$.

It turns out that such a fast rotation cannot be a general result of stellar
evolution. Although young massive star often rotate sufficiently rapidly at
birth, their cores are expected to experience strong spin-down during the red
giant phase and during the intensive mass loss period characteristic for
massive stars at the Wolf-Rayet phase \citep{HWS05}. In fact, this theoretical
result agrees very well with the observed rotation rates of newly born
pulsars.  Thus in order to retain the rotation rate required in the collapsar
model, the evolution of LGRB progenitors must proceed along a rather exotic
route.  Recently, it was proposed that a combination of low metalicity and
extremely fast initial rotation, at around $50\%$ of the break-up speed, could
lead to such a route \citep{YL05,WH06,YLN06}.  On one hand, the mass-loss rate
decreases significantly with metalicity, leading to a significant reduction in
the total loss of angular momentum.  On the other hand, the rotationally
induced circulation becomes very effective at such a high rotation rate and
may result in chemically homogeneous stars that avoid the development of
extended envelops and hence the spin-down of stellar cores via interaction
with these envelopes.  Moreover, the star remains compact by the time of its
collapse so the LGRB jet can break out from the star on the time scale
compatible with the observed durations of LGRBs.

Another exotic scenario involves close high-mass binary systems, where the
fast rotation of stellar cores is sustained via the tidal interaction between
companions \citep{TC03,TC04,I04,P04,HY07}. In this case, the pre-supernova is
a compact helium star, essentially a Wolf-Rayet (WR) star, because the
extended envelope is dispersed into the surrounding space during the common
envelope phase.  The stellar rotation in such systems is synchronised with the
orbital motion on a very short time scale \citep[e.g.][]{HY07}.  The
contraction of CO-cores during stellar evolution leads to their additional
spin-up but due to the core-envelope coupling only a fraction of their angular
momentum is retained \citep{YLN06}. As the result, the core rotation rate is
insufficient in the cases where the companion of the helium star is a
main-sequence star. According to \citet{HY07}, the core rotation can be high
enough to fit the collapsar model with the neutrino-driven LGRB jet only if
the component is also a compact star, namely NS or BH. Three examples of such
systems are known to date: Cyg X-3, IC 10 X-1, and NGC 300 X-1.  Cyg X-3 has a
very short orbital period, only 4.8h \citep{vk92}, and the radius of the WR
star in this system is less than $3-6 R_{\odot}$ \citep{cm94}. The recently
discovered IC 10 X-1 and NGC 300 X-1 have the orbital periods of 35h and 33h
respectively \citep{cp07,pc07,fil08}. The masses of Wolf-Rayet stars are
estimated at $18-40 M_{\odot}$ for NGC 300 X-1 and $\simeq35 M_{\odot}$ for IC
10 X-1 \citep{cc04}.  Given the observed production rate of such systems
\citet{HY07} predicted one hypernova/LGRB every 2000 years in a galaxy similar
to our own.

The neutrino heating is not the only possible mechanism behind the explosions
of collapsing stars.  Perhaps somewhat less popular, but the magnetic
mechanisms are also regarded as potentially important
\citep[e.g.][]{BK70,LW70,MWH01,MBA06,BDLOM07}.  Likewise, the LGRB jets can
also be powered via a magnetic mechanism, in particular the Blandford-Znajek
mechanism, which utilises the rotational energy of the BH
\citep[e.g.][]{BZ77,MR97,LBW00,PMAB03,M06,BK08,KB09}.  Here the black hole is
also required to rotate quite rapidly. However, the efficiency of this
mechanism is not that sensitive to the mass accretion rate and such rapid
rotation does not have to be achieved right after the collapse of the Fe core.
Instead, it can be built up gradually during the rest of the stellar collapse.
This difference in the sensitivity to mass accretion rate favours the BZ
mechanism over the neutrino mechanism in the case of very long-duration LGRBs,
more than 100 seconds long \citep{MWH01}.  The discovery by {\it Swift} of the
shallow decay phase and late flares in the X-ray light curves of LGRBs
\citep{z07,ch07} also suggests that the central engine may remain active for
as long as $10^4$ seconds \citep[e.g.][]{lg07,lg08}.  Since the neutrino
mechanism requires the mass accretion rate to stay above
few$\times10^{-2}M_\odot/$s, such a prolonged activity implies the progenitor
mass in excess of few$\times 10^2M_\odot$, which is highly
unlikely\footnote{Typically, the mass of WR star is 9-25 $M_{\odot}$, though
some observations suggested that it can be as high as 83 $M_{\odot}$
\citep{sch99, crw07}.}.

Another problem for the model of neutrino-driven GRBs are the strong precursors
sometimes observed before the arrival of the main gamma-ray pulse \citep{Bur08}.
According to the analysis of \citet{WM07} such precursor and the main pulse
can be attributed to a single eruptive event only when the precursor and the
main pulse are separated by few seconds.  However, in some GRBs the delay can
be as long as few hundreds of seconds and in such cases it is much more likely
that the precursor and the main pulse correspond to two different events in
the life of the central engine.  They proposed that the precursor is produced
during the supernova explosion, in the jet powered by a rotating magnetised
neutron star, and that the main pulse is produced during the fallback phase
when the neutron star collapses into a black hole\footnote{This model may
struggle to explain delays shorter than the typical fallback time, 100-1000
seconds, found in one-dimensional simulations \citet{MWH01}.  However, due to
the rotational effects the supernova explosions could be highly aspherical,
resulting in shorter fallback time scales in the equatorial region.}.  The
typical mass accretion rates in the fallback scenario,
$10^{-2}-10^{-3}M_\odot\,\mbox{s}^{-1}$, are too low for the neutrino
annihilation mechanism and thus this explanation implies magnetic origin for
the main pulse as well \citep{MWH01}.

Thus, the observations require to include the magnetic mechanism, either in
the black hole or, in fact, in the disk version, or both, in the collapsar
scenario. This widens the range of potential progenitors of LGRBs.  Indeed, we
no longer need to constrain ourself to the stars with extremely rapidly
rotating cores but can also include the cases with slower rotation where the
accretion disk forms much later during the course of stellar collapse.
     
In this paper we re-examine the scenario of binary progenitor of LGRBs
allowing for the late formation of accretion disks and lower mass accretion
rates compared to those required in the collapsar model with the neutrino
mechanism.  In Section \ref{FAD} we determine the parameters of binary systems
which allow formation of accretion disks during the collapse of WR companion.
We also estimate masses and spins of the black holes by the time of accretion
disk formation using simplified analytical model for the structure of
pre-supernovae due to \citet{b90}.  In Section \ref{GBH} we investigate the
degree to which the black hole spin can increase later on, during the disk
accretion phase, using the same approach as in the recent study by
\citet{jmp08}. Here we consider not only the Bethe's model but also the
polytropic model and the models of pre-supernovae based on detailed
calculations of stellar evolution.  In Section~\ref{JS} we describe the
numerical simulations of LGRB jet formation with setup based on the results
obtained in the previous Sections. In Section~\ref{CE} we analyse the
potential of the binary scenario in the extreme case, which involves merger of
the WR star with its ultra-compact companion, BH or NS. In Section~\ref{DC} we
summarise our main results and discuss their astrophysical implications.
   
%Nearly 15\% single WR show evidence of rotation \citep{hhh98} fast enough to
%form accretion disk around even Schwarzschild BHs. The observed binary
%fraction among Milky Way WR stars is 40\% \citep{vdh01}.

%%%%%%%%%%%%%%%%%%%%%%%%%%%%%%%%%%%%%%%%%%%%%%%%%%
\section{Formation of accretion disc}
%%%%%%%%%%%%%%%%%%%%%%%%%%%%%%%%%%%%%%%%%%%%%%%%%%
\label{FAD}

In a synchronised binary the tidal torques force the components to spin with
the same rate as the orbital rotation,
\begin{equation} \Omega^2_s = GM_s(1+q)/L^3,
\label{omega_s}
\end{equation} where $L$ is the orbital separation, $M_s$ is the mass of the
star under consideration, and $q=M_{com}/M_{s}$, where $M_{com}$ is the mass
of the companion star.  Since, the orbital frequency decreases with $L$, the
maximum possible spin is reached when the separation is minimum. This
corresponds to the case where the star radius is about the size of its Roche
lobe. The relation between the minimum separation $L_{min}$, the stellar
radius $R_s$, and $q$ can be approximated with sufficient accuracy for
$1/100<q<100$ as
\begin{equation} L_{min} = 2.64 q^{0.2084} R_s
\label{Lmin}
\end{equation} \citep{pk64}.

During the stellar collapse the centrifugal force will halt the free-fall of
the outer layers and promote the development of accretion disk provided the
specific angular momentum on the stellar equator exceeds that of the
marginally bound circular orbit for the black hole with the same mass and
angular momentum as the star.  The angular momentum of Kerr black holes is
$$
J_{h}=a\frac{GM_h^2}{c},
$$
where $-1 < a < 1$ is the dimensionless spin parameter and $M_h$ is the hole
mass. The specific angular momentum of test massive particles on circular
orbits in the equatorial plane is 
\begin{equation} l=\frac{(r^{2}- 2ar^{1/2}+ a^{2})} {r^{3/4}(r^{3/2}-3r^{1/2}
+2a)^{1/2}} \frac{GM_h}{c},
\label{lmu}
\end{equation} where $r=R/R_g$ and $R_g=GM_h/c^2$, and the radius of the
marginally bound orbit is
\begin{equation} r_{mb} = \lbrace 2 - a + 2(1 - a)^{1/2} \rbrace
\label{rmb}
\end{equation} \citep{bard72}. The disk formation condition is  

\begin{equation} 
\Omega_s R_s^2 > l_{mb},
\end{equation} 
where $l_{mb}=l(r_{mb})$. As we shell see later, at the time of the disk
formation $a$ is quite small. Using the Taylor expansion we find that

\begin{equation} l_{mb} = (4-a)\frac{GM_h}{c} + O(a^2) \simeq \frac{4GM_h}{c}.
\label{rmb_s}
\end{equation} 
Using this result and Eq.\ref{omega_s} we can now write the disk formation 
condition as

\begin{equation} 
\left(\frac{L}{R_s} \right)^3 < \frac{1+q}{16} r_s,
\label{eq2}
\end{equation} 
where $r_s=R_s/R_{gs}$ and $R_{gs}= GM_s/c^2$.  
For the typical parameters of WR stars this amounts to
\begin{equation} L < 14 R_s (1+q)^{1/3} \left(\frac{R_s}{R_\odot}\right)^{1/3}
\left(\frac{M_s}{10M_\odot}\right)^{-1/3}.
\label{eq2a}
\end{equation} 
The comparison of this result with Eq.\ref{Lmin} shows that collapse of WR
stars in close binaries can indeed lead to formation of accretion discs
\citep{bt04,P04}.  We can rewrite the above condition in terms of the binary
period, $T_b$, as

\begin{equation} T_b < \frac{1}{4}T_k r_s^{-1/2} \simeq 48\mbox{hr}
\left(\frac{M_s}{10M_\odot}\right)^{-1} \left(\frac{R_s}{R_\odot}\right)^2,
\label{eq2b}
\end{equation} where $T_k$ is the Keplerian period at $R=R_s$. This upper
limit is about five times higher than that obtained in \citet{P04} who
required the disk to form immediately after the collapse of iron core.

Effectively cooled accretion disks remain geometrically thin and their inner
radius is given by the radius of the last stable circular orbit

\begin{equation} 
r_{ms} = \lbrace 3+Z_2 - [(3-Z_1)(3+Z_1+2Z_2)]^{1/2} \rbrace,
\label{rms}
\end{equation} 
where $r_{ms} = R_{ms}/R_g$ and

\begin{equation}
\begin{array}{l} Z_1 \equiv 1+(1-a^2)^{1/3}[(1+a)^{1/3}+(1-a)^{1/3}],\\ Z_2
\equiv (3a^2+Z_1^2)^{1/2} .
\end{array}
\end{equation} \citep{bard72}.  
The corresponding specific angular momentum,
$l_{ms}$, determines the evolution of the black hole spin via the disk
accretion\footnote{For simplicity, we ignore the effects of magnetic torques
on the evolution of black hole spin.}.

%fffffffffffffffffffffffffffffffffffffffffffffffffffffffffffffffff
\begin{figure}
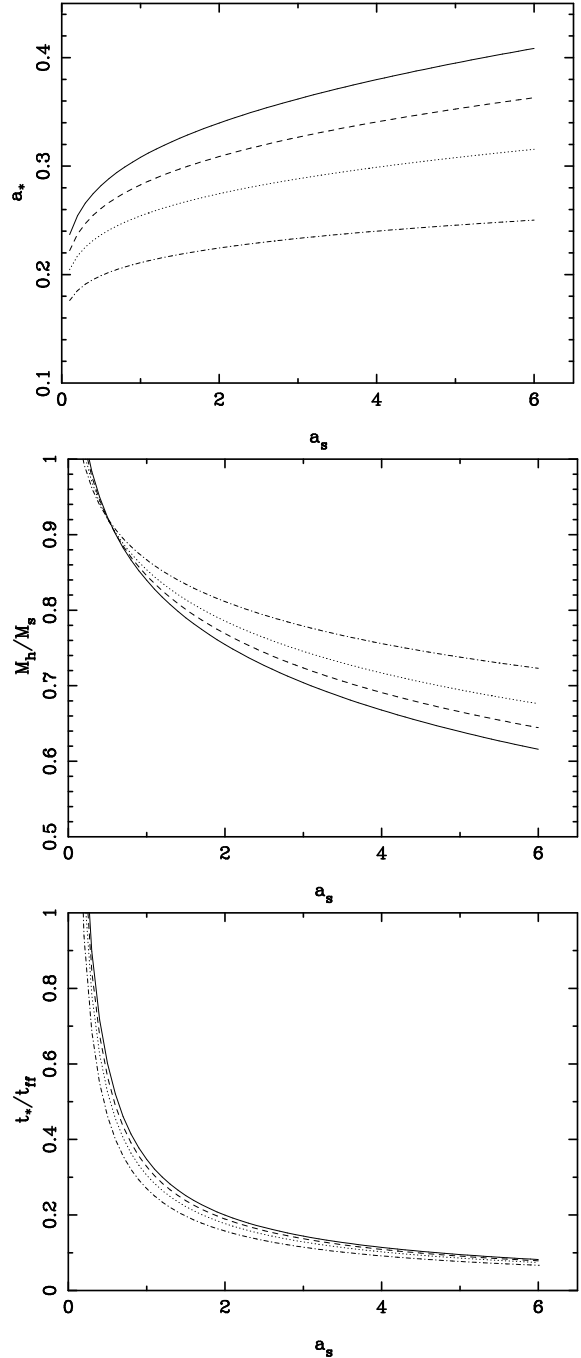

\includegraphics[width=60mm,angle=-90]{figures/fig1a.eps}
\includegraphics[width=60mm,angle=-90]{figures/fig1b.eps}
\includegraphics[width=60mm,angle=-90]{figures/fig1c.eps}
\caption{The black hole spin (top panel) and mass (middle panel) at the
disk formation time as functions of the progenitor spin, $a_s$, 
for $M_c/M_s=1/3$ (dash-dotted line), 1/5 (dotted line), 1/9 (dashed line),
and 1/31 (solid line). The bottom panel shows the time of disk formation as 
a function of $a_s$ for the same models.
}
\label{disk_formed}
\end{figure}
%fffffffffffffffffffffffffffffffffffffffffffffffffffffffffffffffff

%fffffffffffffffffffffffffffffffffffffffffffffffffffffffffffffffff
\begin{figure*}
\includegraphics[width=84mm,angle=-0]{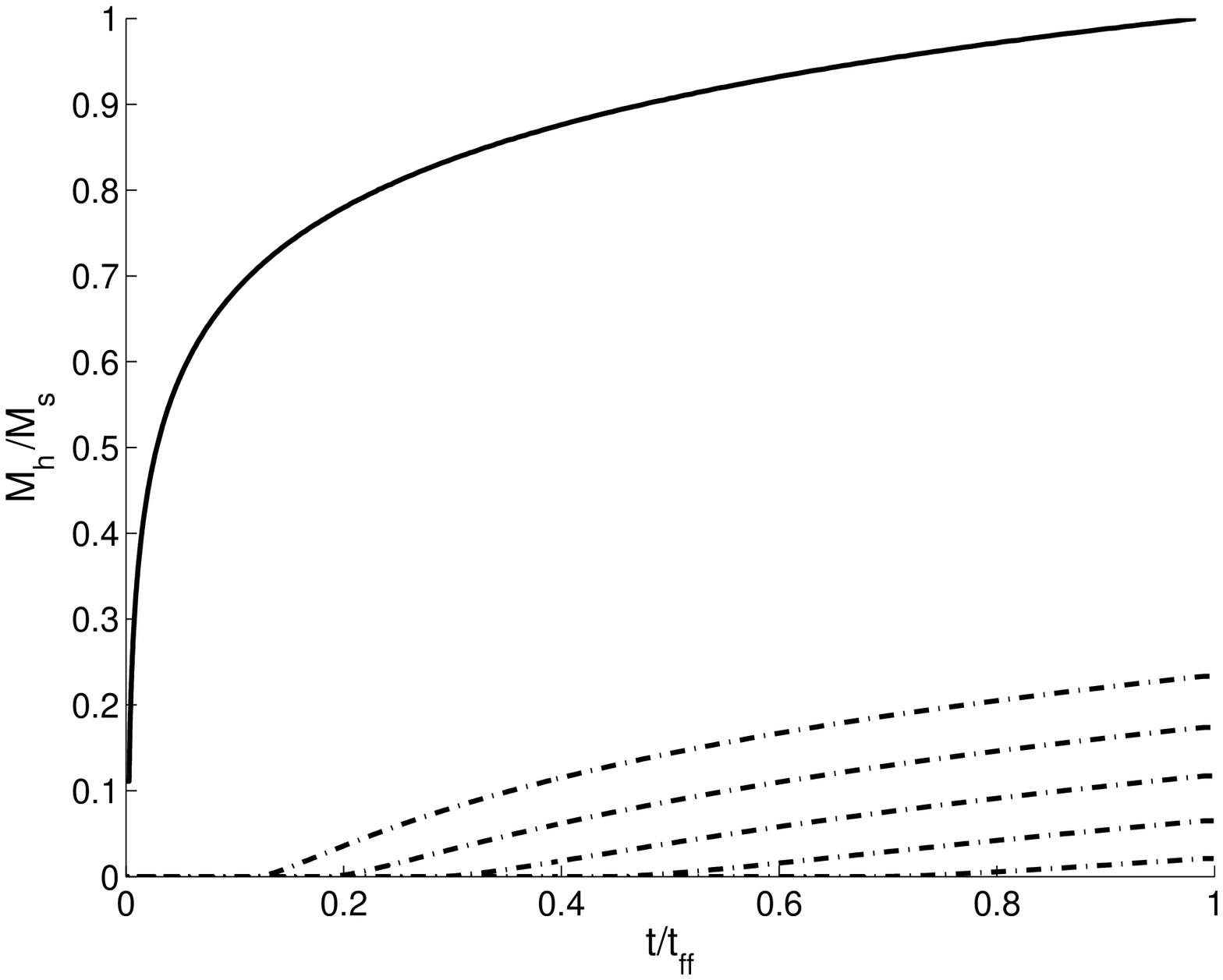}
\includegraphics[width=84mm,angle=-0]{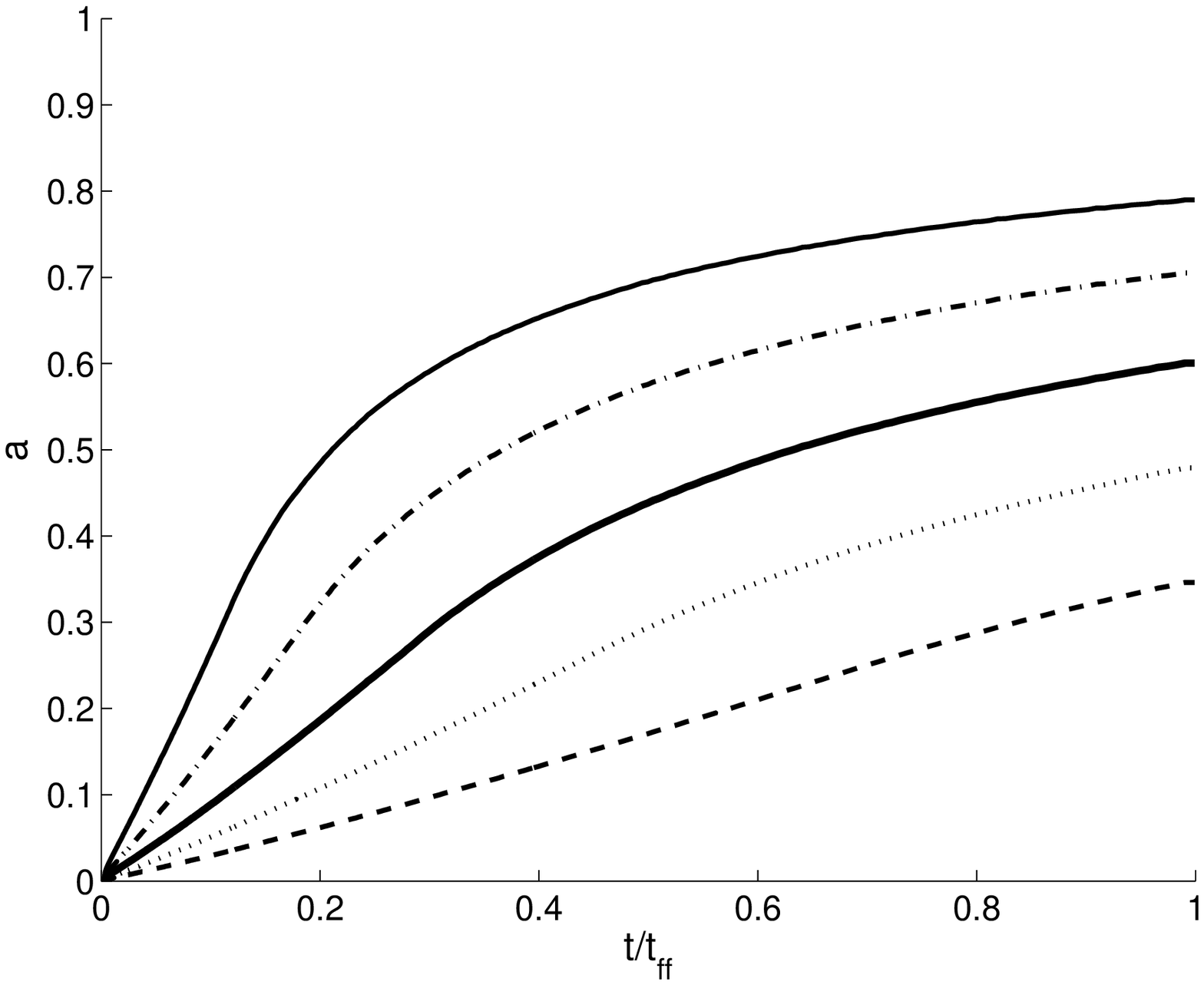}
\caption{ Evolution of the black hole mass and spin in the Bethe's model with
$M_c/M_s=1/9$.  {\it Left panel:} The total mass of black hole (solid line)
and the mass accumulated via the accretion disk (dash-dotted lines) for the
progenitor spin $a_s=0.33, 0.58, 1.0, 1.7$, and 3. The higher value of $a_s$
correspond to the higher fraction of mass processed via the disk.  {\it Right
panel:} The spin parameter $a$ of the black hole for the progenitor spin
$a_s=0.33$ (dashed line), 0.58 (dotted line), 1.0 (thick solid line), 1.7
(dot-dashed line), and 3.0 (thin solid line). The evolution time is given in
the units of the free-fall time (see Eq.\ref{te}).}
\label{bethe-ma}
\end{figure*}
%fffffffffffffffffffffffffffffffffffffffffffffffffffffffffffffffff

The outer radius of the disk, $R_d$, is determined by the specific angular
momentum on the stellar equator\footnote{In fact, various torques operating in
the accretion disk change the angular momentum and hence the location of the
outer edge but this effect is relatively minor \citep{shs73,CB07}.},
$l_s=\Omega_s R_s^2$.  Assuming that $r_d=R_d/R_{gs} \gg 1$ the angular
momentum at the outer edge of the disk is simply 
$$
l_d=(GM_sR_d)^{-1/2}.
$$ 
Matching $l_d$ and $l_s$, and using Eq.\ref{omega_s}, we find that

\begin{equation} 
r_d = r_s (1+q) \left(\frac{L}{R_s}\right)^{-3}.
\end{equation} 
Thus,

\begin{equation} 
r_d \sim 47 \left(\frac{R_s}{R_\odot}\right)
\left(\frac{M_s}{10M_\odot}\right)^{-1} \left(\frac{\tilde{L}}{10}\right)^{-3}
\end{equation} 
where $\tilde{L}=(1+q)^{-1/3}(L/R_s)$. For the widest orbital
separation which still allows disk formation (see Eq.\ref{eq2a}) this equation
gives $r_d\le 17$ whereas for the closest one (see Eq.\ref{Lmin}) we have
$r_d\le 5\times10^3$.  Thus, the model predicts a wide range of accretion disk
sizes.  Compact disks and the inner regions of large disks will cool via the
neutrino emission, whereas the outer regions of large disks will remain
adiabatic. The accretion time of neutrino cooled disks

\begin{equation} t_{d} \approx 2.6 \left(\frac{\alpha}{0.1}\right)^{-6/5}
\left(\frac{r_d}{100}\right)^{4/5}\left(\frac{M_h}{10M_{\odot}}\right)^{6/5}
\mbox{s}
\label{tdn}
\end{equation} \citep{pwf99} is significantly less than the free-fall time
scale

\begin{equation} t_{f\!f} \approx 240
\left(\frac{R}{R_{\odot}}\right)^{3/2}\left(\frac{M_s}{10M_{\odot}}\right)^{-1/2}
\mbox{s}.
\label{te}
\end{equation} The accretion time of large disks can be estimated using the
$\alpha$-model for slim disks ($\delta = H_d/R_d \simeq 0.3$)

\begin{equation} t_{d} \approx 250
\left(\frac{\alpha\delta^2}{0.01}\right)^{-1}
\left(\frac{r_d}{10^3}\right)^{3/2}\left(\frac{M_h}{10M_{\odot}}\right)
\mbox{s}
\label{tdn_ss}
\end{equation} \citep{shs73}.  Thus, with the exception of largest disks, the
time scale of disk accretion is shorter compared to the free-fall time scale,
and hence the growth rate of the black hole mass is given directly by the 
rate of the collapse.

In order to estimate the mass and rotation rate of the black hole at the time
of the disk formation one needs to know the mass distribution of progenitor at
the onset of collapse. Here we adopt the power law model used by \citet{b90}
in his analytical models of core-collapse supernovae,
\begin{equation} \rho(R)=\rho_c \left(\frac{R}{R_c}\right)^{-3}, \qquad R>R_c,
\end{equation} where $R_c$ is the radius of iron core. Simple integration
allows us to find the following equations for the mass

\begin{equation} M(R)=4\pi\rho_c R_c^3 \ln(R/R_c)
\label{M}
\end{equation} and the moment of inertia

\begin{equation} I(R)=\frac{1}{3}\frac{M(R)R^2}{\ln(R/R_c)}
\label{I}
\end{equation} of the shell between the iron core and the radius $R$.

By analogy with the black hole theory it is convenient to describe the
rotation rate of collapsing star using the spin parameter
\begin{equation} a_s=\frac{J_s c}{GM_s^2}.
\label{a}
\end{equation} In Bethe's model it relates to $\Omega_s$ via

\begin{equation} \Omega_s= a_s \frac{3GM_s (1+\eta)^2}{R_s^2 c} \ln y_s
\label{a-omega}
\end{equation} where $y_s=R_s/R_c$, $\eta=M_c/M_s$, and we ignore the small
contribution of compact iron core to the total spin of the star.  The
condition (\ref{Lmin}) with $q=1$ implies that

\begin{equation} a_s \leq \frac{1}{9\ln y_s} r_s^{1/2} \simeq 5.2
\fracb{R}{R_\odot}^{1/2} \fracb{M_s}{10M_\odot}^{-1/2},
\label{as_max}
\end{equation} where we used $y_s=100$. This seems to suggest that the 
stellar collapse may lead to formation of rapidly rotating black holes. 

Suppose that the disk is first formed at time $t^*$ and that by this time the
black hole has swallowed the star up to the initial radius $R=R_*$.  Assuming
that the black hole spin at this point is low, $a_*\ll 1$, we have
\begin{equation} \Omega_s R_*^2 =(4-a_*) \frac{G(M_*+\eta M_s)}{c},
\label{cond2}
\end{equation} where $M_*=M(R_*)$.  Using Eqs.(\ref{M}) and (\ref{a-omega})
this condition can be written as the following algebraic equation for
$y_*=R_*/R_c$
  
\begin{equation} y_*^2 = \frac{(4-a_*)y_s^2}{3 (1+\eta)^2 a_s \ln y_s} \left[
\frac{\ln y_*}{\ln y_s}+\eta \right],
\end{equation} where
\begin{equation} a_* = \frac{4}{1+3(\ln y_*+\eta\ln y_s)}.
\end{equation} This equation is solved numerically and the results are
presented in Fig.\ref{disk_formed}.  One can see that the disk is formed
relatively late, with the typical time $t_*>0.1t_{f\!f}$, when more than a
half of the star has already collapsed into the black hole.  However, the black 
hole spin at this moment is relatively low, $0.2<a_*<0.4$.

%fffffffffffffffffffffffffffffffffffffffffffffffffffffffffffffffff
\begin{figure}
\includegraphics[width=84mm,angle=-0]{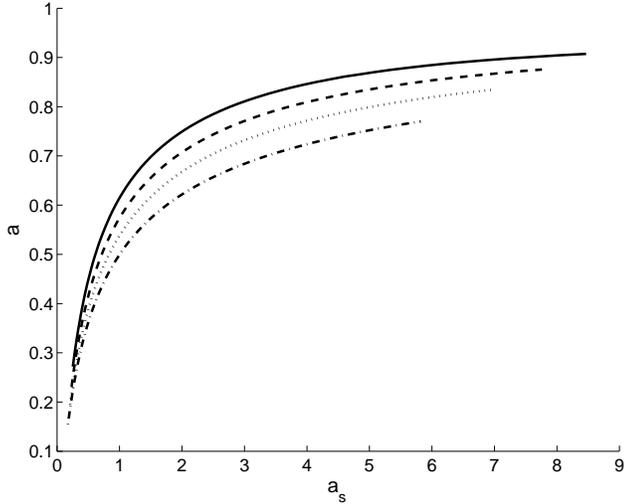}
\caption{The final value of the black hole spin in Bethe's model with as a
function of the progenitor spin for models with $M_c/M_{star}=1/3$ (dot-dashed
line), 1/5 (dotted line), $1/9$ (dashed line), and $1/31$ (solid line). }
\label{bethe-a}
\end{figure}
%fffffffffffffffffffffffffffffffffffffffffffffffffffffffffffffffff

%%%%%%%%%%%%%%%%%%%%%%%%%%%%%%%%%%%%%%%%%%%%%%%%%%
\section{Growth of Black Holes}
%%%%%%%%%%%%%%%%%%%%%%%%%%%%%%%%%%%%%%%%%%%%%%%%%%
\label{GBH}

In order to explore the evolution of the black hole spin for more
sophisticated models of LGRB progenitors, as well as its evolution in Bethe's
model after the disk formation, one can integrate the following system of
dynamic equations

\begin{equation} \frac{dM_h}{dR} = 4\pi R^2 \rho(R),
\label{M-evolv}
\end{equation}
 
\begin{equation} \frac{dJ_h}{dR} = 4\pi R^2 \rho(R) \int\limits_0^{\pi/2}
\tilde{l}(R,\theta) \sin{\theta} d\theta .
\label{J-evolv}
\end{equation} Here, $\rho(R)$ is the stellar mass density prior to the
collapse and $\tilde{l}(R,\theta)$ is the specific angular momentum retained
by the fluid element, initially located at the point with the coordinates
$\{R,\theta\}$, by the time it crosses the event horizon. 
This quantity is given by

\begin{equation} \tilde{l}=\left\{ \begin{array}{lcl} l(R,\theta) &\text{if}&
l<l_{mb}(M_h,J_h)\\ l_{ms}(M_h,J_h) &\text{if}& l>l_{mb}(M_h,J_h)
\end{array} \right. ,
\end{equation} 
where $l(R,\theta)$ is the distribution of the progenitor's
angular momentum.  The initial conditions for
Eqs.(\ref{M-evolv},\ref{J-evolv}) correspond to the iron core of the WR star

\begin{equation} 
M_h(R_c)=M_c,\quad J_h(R_c)=0,
\end{equation} 
where $M_c$ and $R_c$ are respectively the mass and the radius
of the core.
When the accretion rate is determined by the free-fall time, $R$ and $t$ can be 
related via 
\begin{equation} 
t^2 = \frac{2R^3}{9GM(R)}.
\label{t_f}
\end{equation} 

%fffffffffffffffffffffffffffffffffffffffffffffffffffffffffffffffff
\begin{figure*}
\includegraphics[width=84mm,angle=-0]{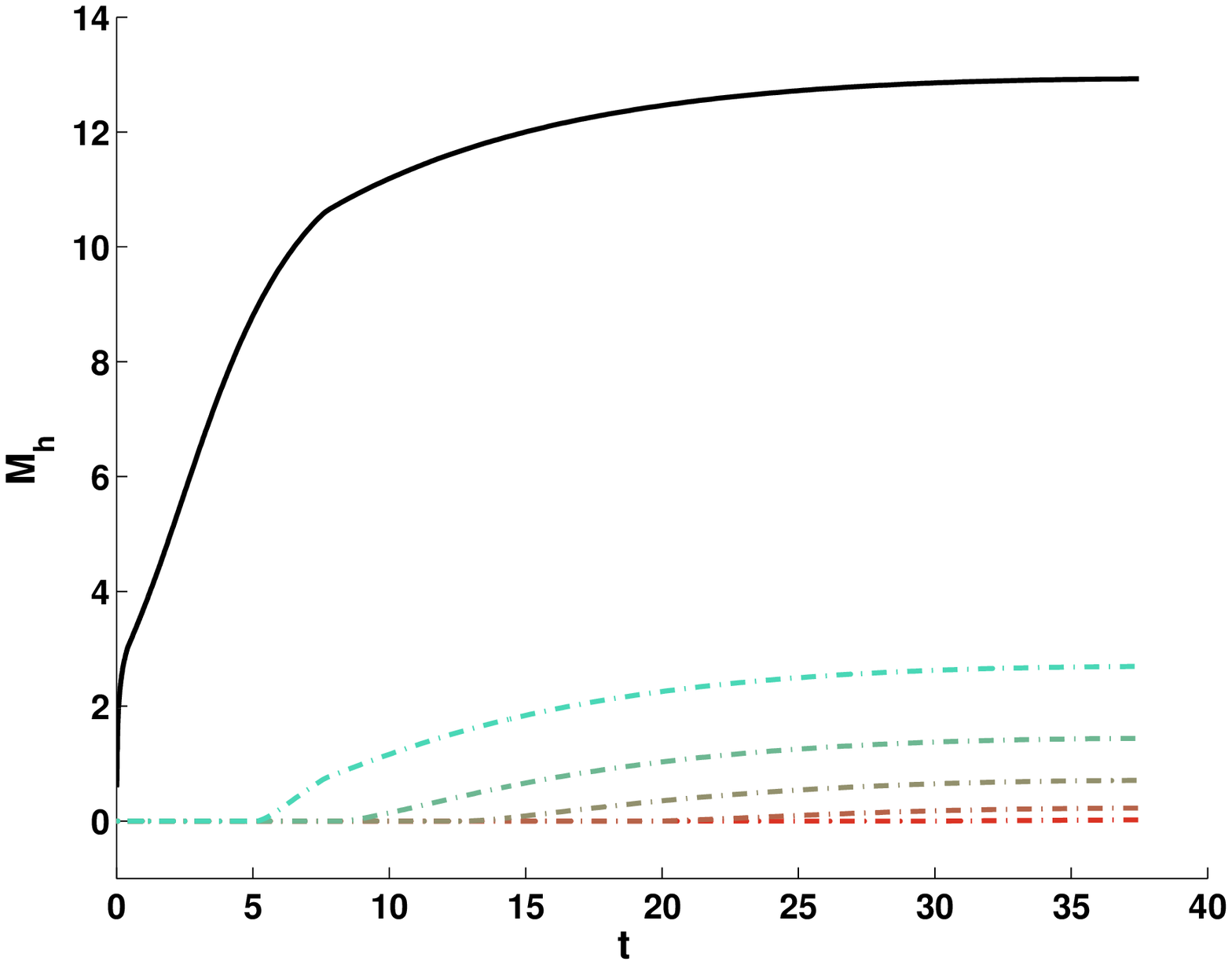}
\includegraphics[width=84mm,angle=-0]{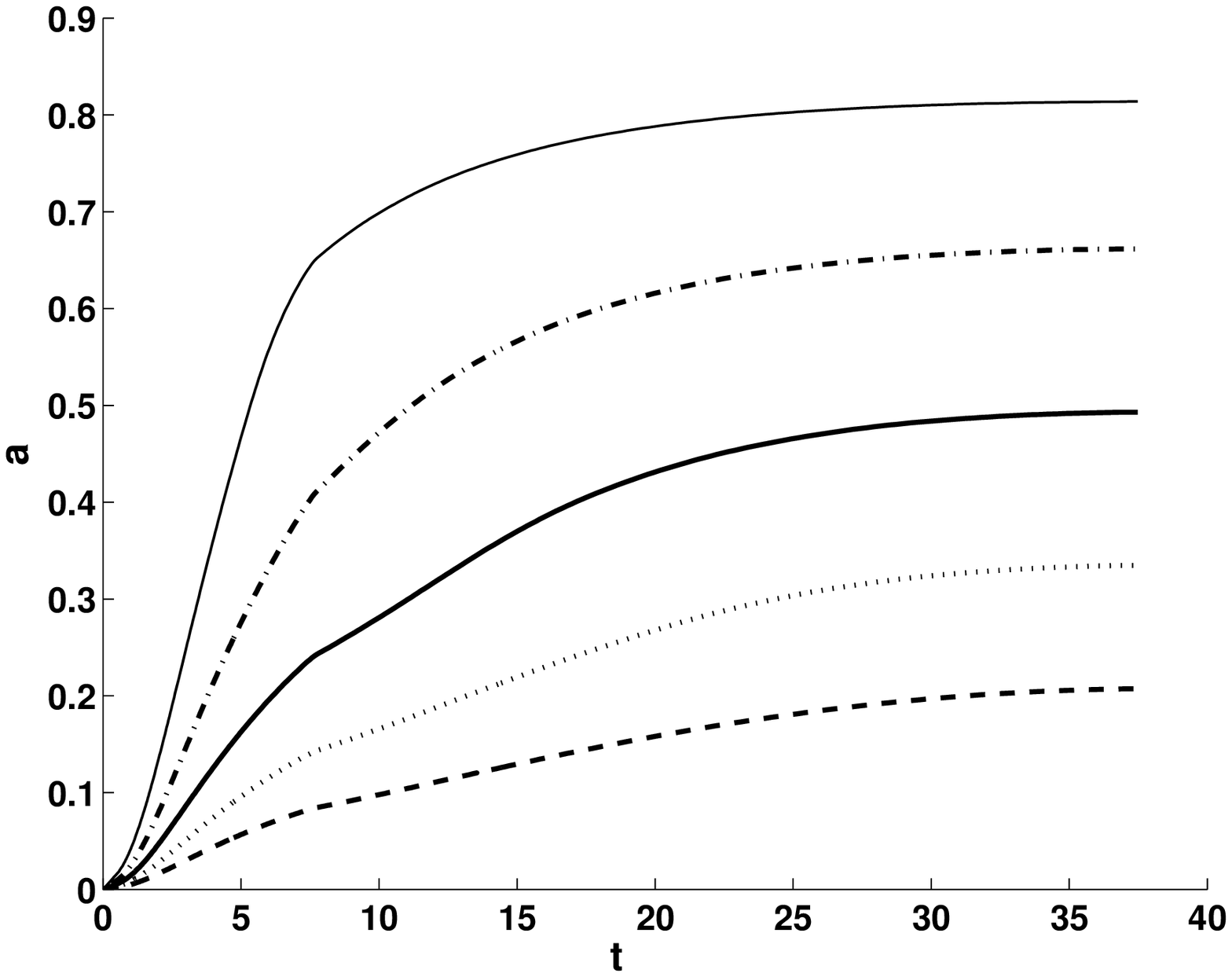}
\caption{ Evolution of black hole's mass and spin in numerical model B.  {\it
Left panel:} The total mass of black hole (solid line) and the mass
accumulated via the accretion disk (dash-dotted lines) for the progenitor spin
$a_s=0.20,0.33,0.57, 0.96$, and 1.6. The higher value of $a_s$ correspond to
the higher fraction of mass processed via the disk.  {\it Right panel:} The
spin parameter $a$ of the black hole for the progenitor spin $a_s=0.20$
(dashed line), 0.33 (dotted line), 0.57 (thick solid line), 0.96 (dot-dashed
line), and 1.6 (thin solid line)}
\label{mod_B_ma}
\end{figure*}
%fffffffffffffffffffffffffffffffffffffffffffffffffffffffffffffffff
%fffffffffffffffffffffffffffffffffffffffffffffffffffffffffffffffff
\begin{figure*}
\includegraphics[width=84mm,angle=-0]{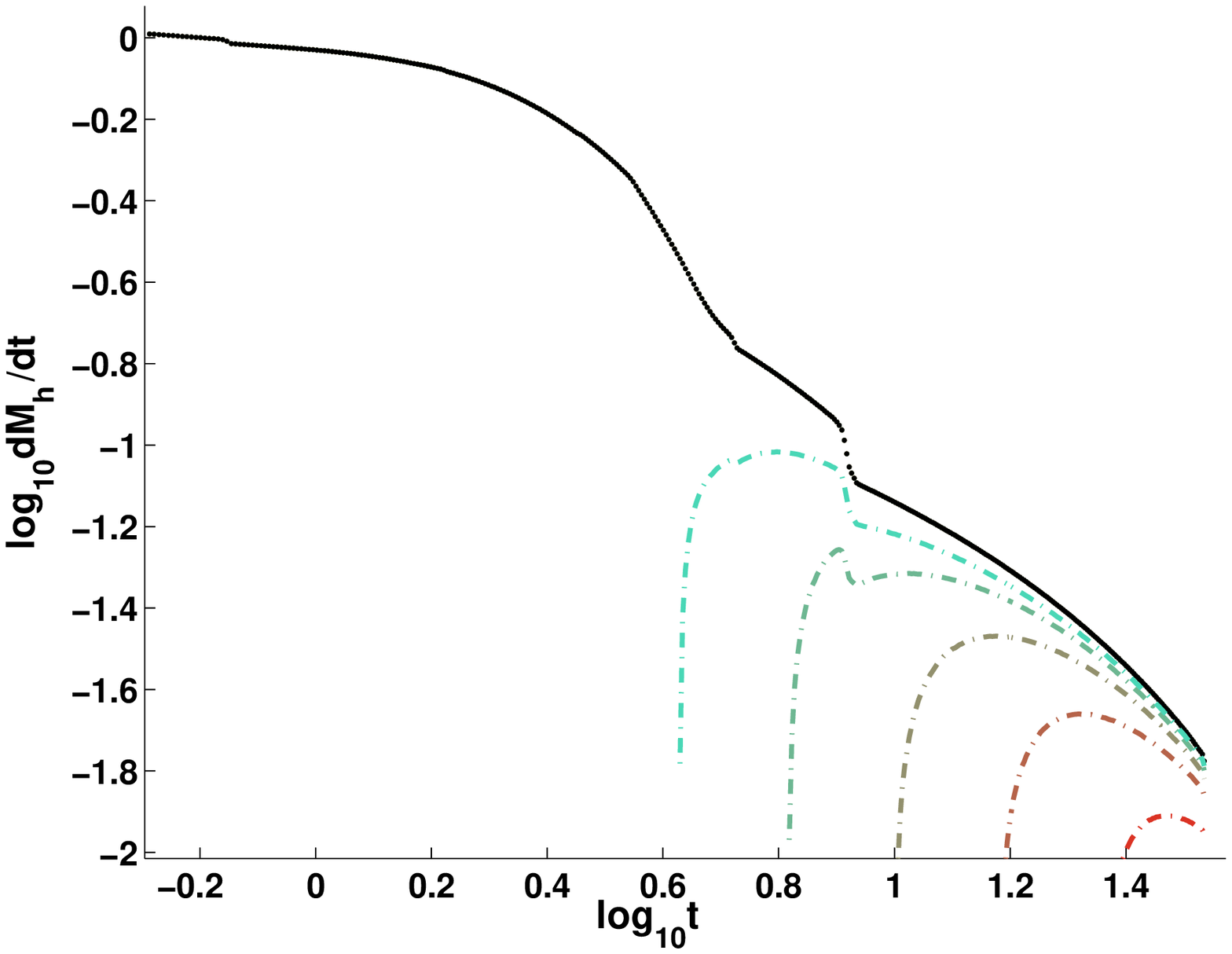}
\includegraphics[width=84mm,angle=-0]{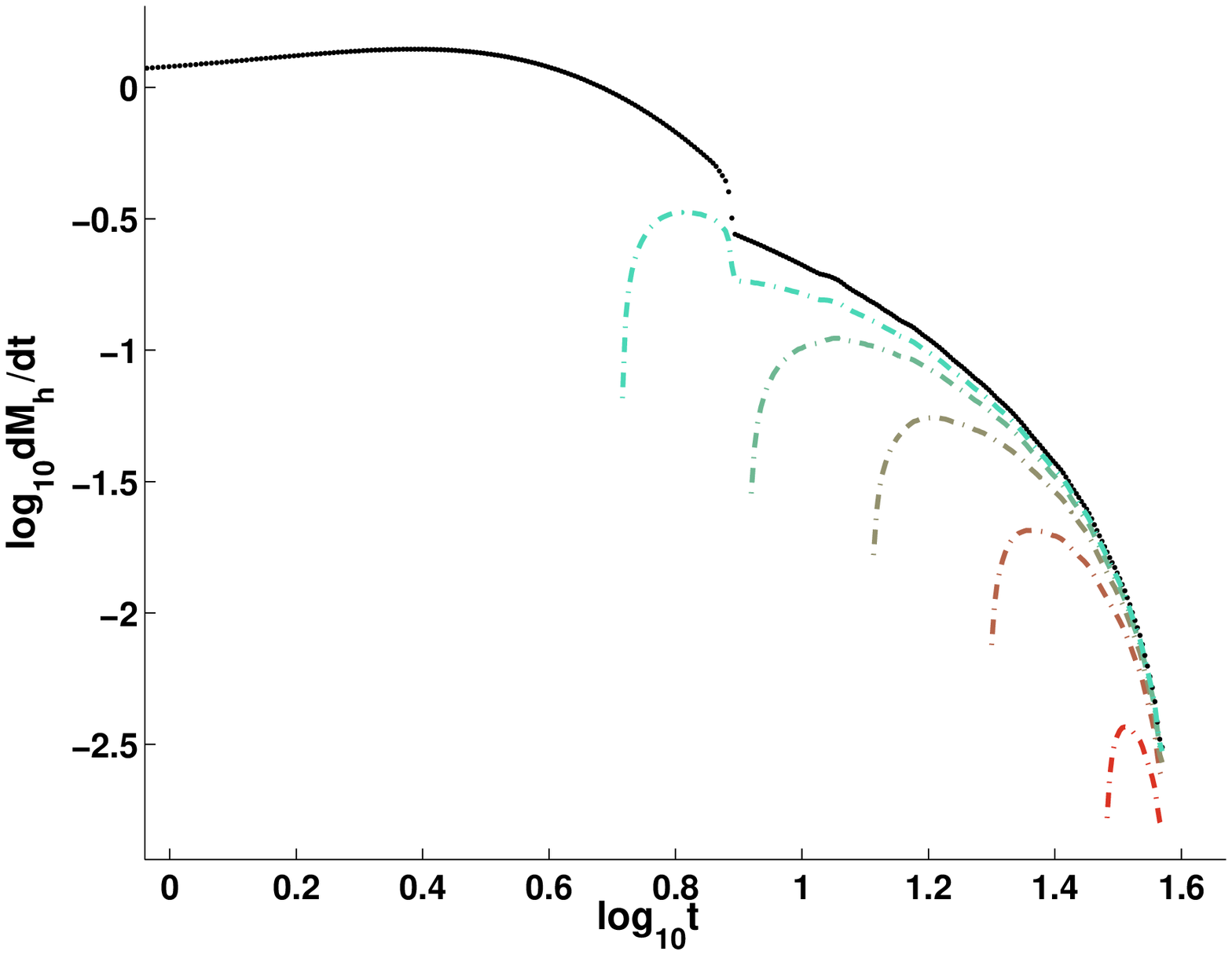}
\caption{The accretion rate (in the units of $M_{\odot} \mbox{ s}^{-1}$) for
model A (left panel) and model B (right panel). The solid lines show the total
accretion rate whereas the dash-dotted lines show the disk accretion rates for
different spins of the progenitor; $a_s=0.32$, 0.54, 0.90, 1.52, and 2.55 for
model A and $a_s=0.20$, 0.33, 0.57, 0.96, and 1.6 for model B. Higher values
of $a_s$ correspond to earlier formation of accretion disk and higher disk
accretion rates.}
\label{mod_AB_mdot}
\end{figure*}
%fffffffffffffffffffffffffffffffffffffffffffffffffffffffffffffffff

The same approach has been used in \citet{jmp08} in their search for the laws
of rotation that would fit the collapsar model of LGRBs. They did not consider
the solid body rotation\footnote{ The solid body rotation law was studied in
\cite{jp08} but it was assumed there that the black hole was non-rotating.}
and assumed that initially the black hole is rapidly rotating, with
$a=0.85$. They also used the model of geometrically thick and radiatively
inefficient disk, with the inner edge located at the radius of the marginally
bound orbit, whereas we use the thin disk approximation, which is more suitable 
for the neutrino-cooled collapsar disks.

%ssssssssssssssssssssssssssssssss
\subsection{Bethe's model}
%ssssssssssssssssssssssssssssssss
\label{bethe}

Figure \ref{bethe-ma} shows the typical evolution of the black hole mass 
and spin, as described by Eqs.\ref{M-evolv} and \ref{J-evolv}, for 
the Bethe's model.  One can
see that the black hole spin increases significantly above the values attained 
by the time of disk formation.  Eventually, it reaches the relatively high values of
$a=0.3-0.8$, the final spin depending mainly on the progenitor spin and less
so on the mass fraction of the iron core (see Fig.\ref{bethe-a}). These higher
values of $a$ imply higher potential efficiency of both the neutrino
annihilation and the Blandford-Znajek mechanisms of the LGRB jet production
\citep{pwf99,ZB09,BK08}.

The total mass accretion rate can be easily derived from the mass
distribution and the free fall time (see Eq.\ref{t_f}):
\begin{equation} \dot{M} = \frac{2}{3}\frac{M_s}{\ln y_s} t^{-1} \simeq 1.45
\fracb{M_s}{10M_\odot} \fracb{t}{1\mbox{s}}^{-1} M_\odot \mbox{s}^{-1},
\label{M-dot}
\end{equation} 
where $t$ is the time since the start of the collapse. As one can see in 
Figure \ref{bethe-ma}, soon after the disk formation the mass 
accretion rate becomes dominated by disk. Initially, the rate can 
be rather high but at around $t\simeq 100$s it becomes insufficient for the 
neutrino annihilation mechanism to operate.

%sssssssssssssssssssssssssssssssssss
\subsection{Stellar evolution models}
\label{SEM} %sssssssssssssssssssssssssssssssssss

Although the Bethe's model provides a reasonable zero-order approximation for
the structure of pre-supernova stars, the more sophisticated models based on
numerical integration of the equations of stellar evolution yield somewhat
different stellar structure with wealth of finer details. Our next results are
based on the pre-collapse structure of massive zero age main sequence (ZAMS)
stars with masses $M_s=20 M_{\odot}$ and $35 M_{\odot}$ described in
\cite{hwls04}.  Assuming that stars of close binaries lose their
extended envelopes we cut of the mass distributions beyond the C/O core.  This
results in the progenitors with masses $M_s=6.15 M_{\odot}$ ( model A ) and
$M_s=12.88 M_{\odot}$ ( model B ) respectively, and radius $R_s\simeq 0.3
R_\odot$. \footnote{ This radius is rather small, twice as small compared to
the models of WR stars constructed in \cite{sm92} and 10-20 time smaller
compared to the observed radii \citep{cm94,crw07}. We can offer no clear
explanation for this discrepancy. Perhaps, the artificial ``removing'' of
extended H/He envelope is not a particularly accurate procedure.} The moments
of inertia of models A and B are $I\simeq 0.065 M_s R_s^2$ and $0.074 M_s
R_s^2$ respectively. Given these parameters, Eqs.\ref{omega_s} and \ref{Lmin}
imply the spin parameters $a_s<2.6$ and $a_s<1.7$ for the models A and B
respectively; somewhat smaller than in the Bethe's model.

Figure \ref{mod_B_ma} shows the evolution of the black hole's mass and spin in
model B for different assumed values of the progenitor's spin. The comparison
with the results obtained for Bethe's model shows only relatively minor
differences, suggesting that Bethe's model is quite accurate. Figure
\ref{mod_AB_mdot} shows the accretions rates, both for the disk and in total,
for different progenitor spins in models A and B. One can see that initially
the disk accretion rate grows rapidly and soon it accounts for most of the
total accretion rate.  Then it begins to decay, approximately as $t^{-1}$ in
model A and $t^{-3}$ in Model B.  For the cases with faster stellar rotation,
the peak disk accretion rate is sufficiently high to ensure effective neutrino
cooling of the disk \citep{CB07}.

%fffffffffffffffffffffffffffffffffffffffffffffffffffffffffffffffff
\begin{figure*}
\includegraphics[width=77mm,angle=-0]{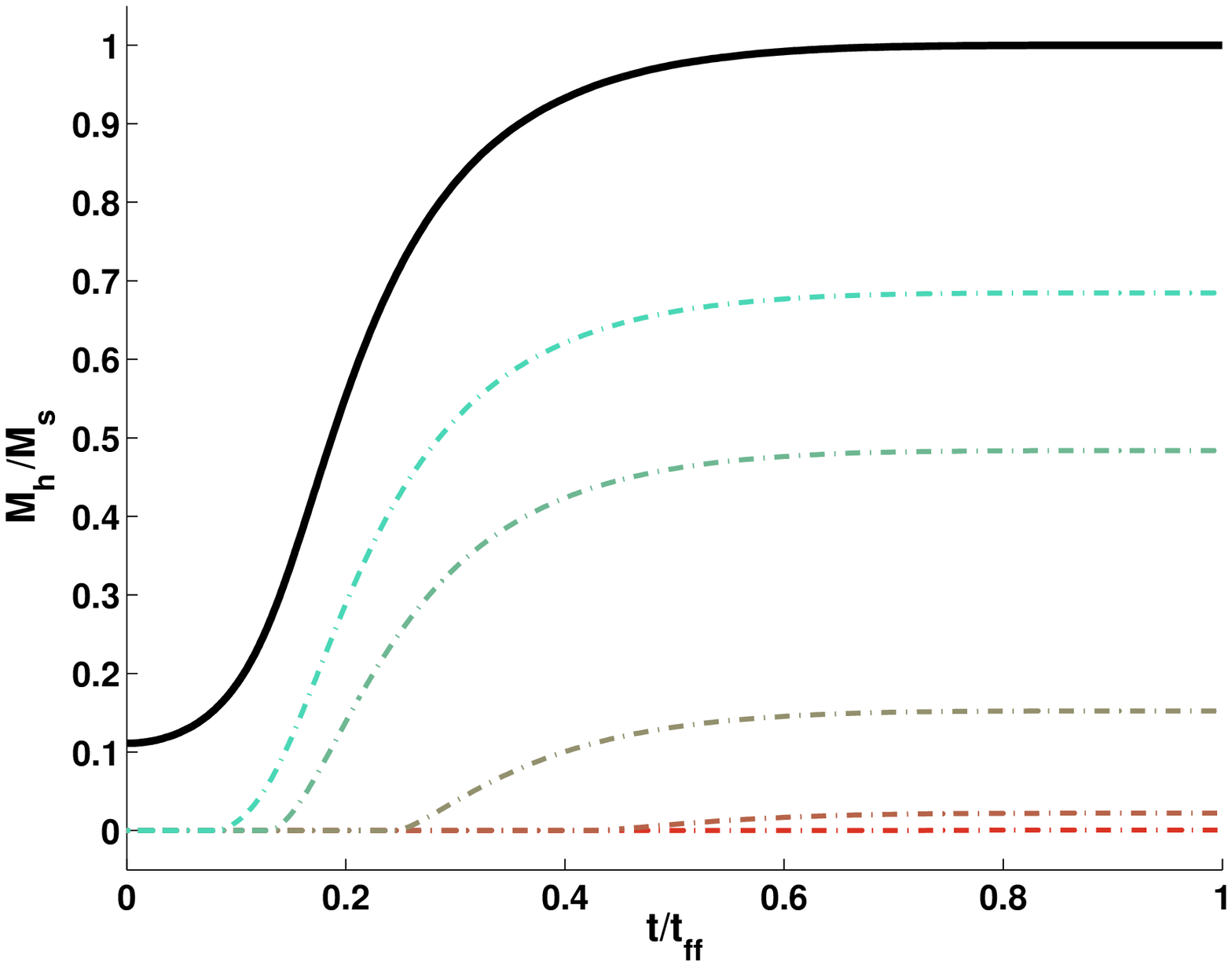}
\includegraphics[width=77mm,angle=-0]{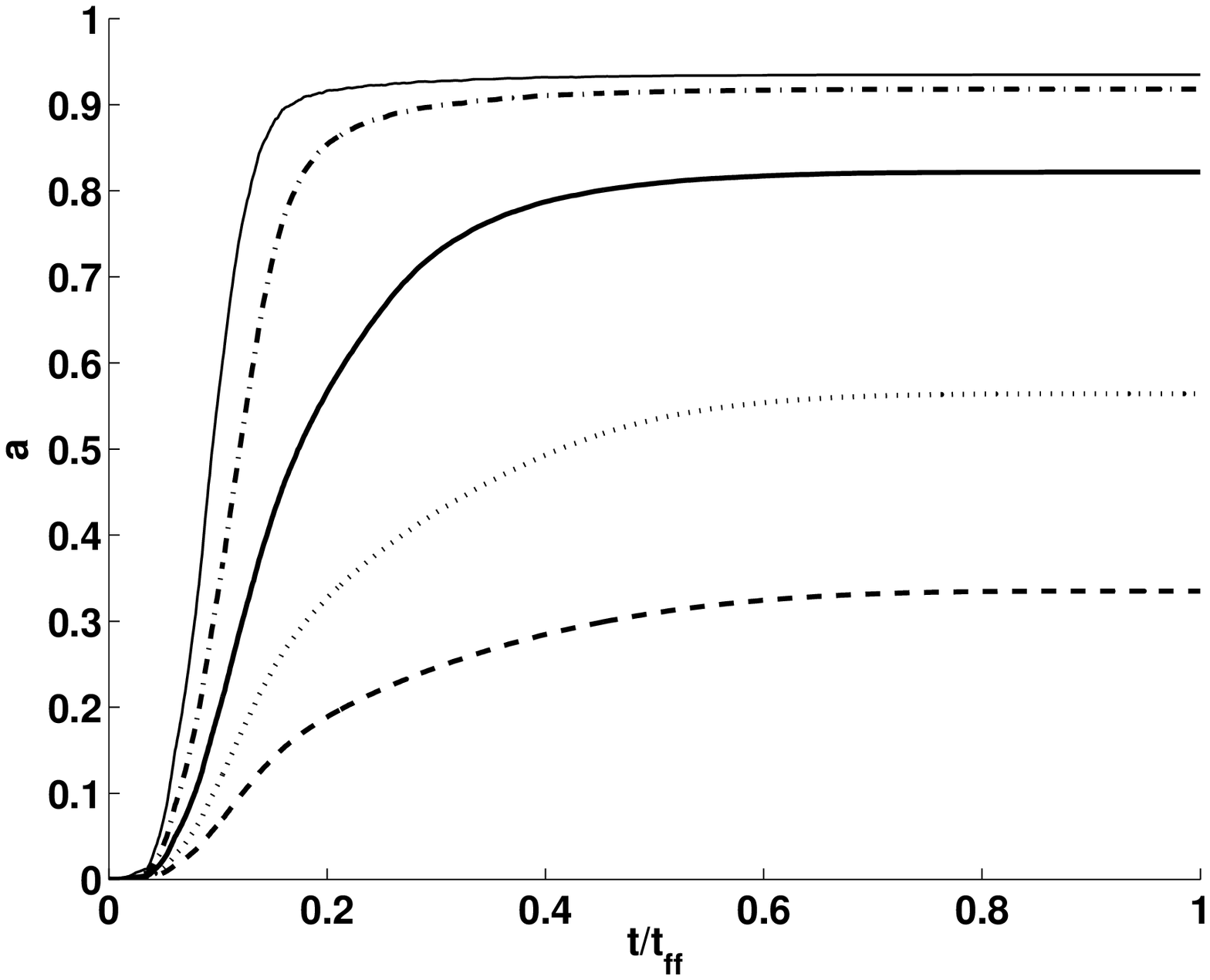}
\caption{ The evolution of black hole's mass and spin in the polytrope model
of progenitor with the initial black hole mass $M_c=M_s/9$. The left panel
shows the total mass of the black hole (thick solid line) as well as the mass
accumulated via the accretion disk for different rotation rates of the
progenitor, $a_s=J_sc/GM_s^2=0.33,\,0.58,\,1.0,\,1.7,\,3.0$ (dash-dotted
lines).  Faster rising lines correspond to higher rotation rate.  The right
panel shows the spin parameter of the black hole, $a$, for the same values of
$a_s$. }
\label{pol_ma}
\end{figure*}
%fffffffffffffffffffffffffffffffffffffffffffffffffffffffffffffffff
 
%sssssssssssssssssssssssssss
\subsection{Polytrope model} 
%sssssssssssssssssssssssssss

\begin{figure}
\includegraphics[width=77mm,angle=-0]{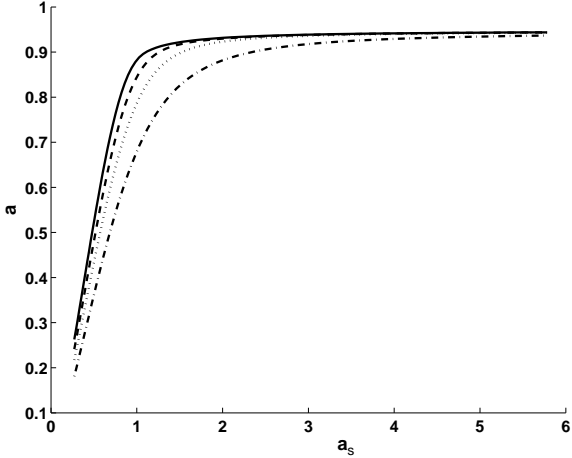}
\caption{The evolution of black hole spin for polytropic models with $a_s=3$
and $M_c/M_{s}=1/3$ (dot-dashed line), 1/5 (dotted line), 1/9 (dashed line),
and 1/31 (solid line). }
\label{pol_fa}
\end{figure}

Finally, we consider the model polytrope with index $n=3$, which could be used to
describe the cores of most massive stars at the pre-supernova phase
\citep{TF07}.  In this model the concentration of mass towards the centre is
much weaker, resulting in higher moment of inertia and larger angular momentum
compared to the Bethe's model with the same mass, radius and rotation
frequency.  Even if we consider models with the same spin parameter $a_s$, the
polytrope yields generally higher fraction of mass accreted via the accretion
disk and more rapidly rotating black holes (see Figure~\ref{pol_ma}).
Similarly to other models, the final value of the black hole's spin 
does not show strong dependence on the iron core mass fraction, at least for
$M_c/M_s \in (1/3,1/31)$ (see Figure \ref{pol_fa}).

The polytrope model was also used to test our calculations against the fully
general relativistic simulations by \cite{shib02}. For the polytropic star
with angular momentum $a_s=1$ our model gives a black hole with $M_b=0.90 M_s$
and $a=0.76$ by the time of disk formation. This is in excellent agreement
with the numerical simulations which give $M_b=0.90$ and $a=0.75$.

%fffffffffffffffffffffffffffffffffffffffffffffffffffffffffffffffff
%fffffffffffffffffffffffffffffffffffffffffffffffffffffffffffffffff

%%%%%%%%%%%%%%%%%%%%%%%%%%%%%%%%%%%%%%%%%%%%%%%%%%
\section{Jet simulations}
%%%%%%%%%%%%%%%%%%%%%%%%%%%%%%%%%%%%%%%%%%%%%%%%%%
\label{JS}

The analysis carried out in Sections \ref{FAD} and \ref{GBH} suggests that
during the collapse of a WR star in a very close binary system, the conditions
can become favourable to production of LGRB jets either via the neutrino
heating or the Blandford-Znajek mechanism. Although the production of jets via
the Blandford-Znajek mechanism has already been studied numerically in several
previous papers the conditions suggested by the binary scenario are different
from those explored so far. By the time of the accretion disk formation the
black hole is much more massive compared to the usually assumed $M_h\simeq
2M_\odot$.  Its rotation rate is noticeably lower compared to $a\simeq 1$, assumed
in the past. Finally, the progenitor's rotation is not differential but
uniform.  These differences invite additional numerical simulations to explore
the new region of parameter space.

%ssssssssssssssssssssssssssss
\subsection{Setup of Simulations} 
%ssssssssssssssssssssssssssss

The progenitor model describes a compact WR star of radius $R_s=3\times
10^{10}$cm and rotation period $T_s=1.4$hr; the corresponding specific angular
momentum on the stellar equator is
$l_s=1.13\times10^{18}\mbox{cm}^2\mbox{s}^{-1}$.  The progenitor's magnetic
field is assumed to be purely poloidal and uniform, with the strength
$B_0=1.4-8.4\times 10^7$G. 

Simulations of this type are computationally expensive even in 2D. 
On the other hand, the early stages of the collapse are
very simple and can be treated analytically with sufficient accuracy. For
these reasons, we start simulations only after the expected time of the disk
formation, $t_s=17$s.  Based on the analysis given in the previous sections,
the black hole mass is set to $M_h=10 M_\odot$ and the mass accretion rate to
$0.14M_{\sun}\mbox{s}^{-1}$. The initial radial distributions of mass and 
velocity are the same as in the Bethe model:

\begin{equation} 
\rho \propto R^{-3/2},\quad v^r=(2GM_h/R)^{1/2}.
\label{ini_v_rho}
\end{equation} 
The initial distributions of angular momentum and magnetic field are 
derived from the progenitor distributions by taking into account the distortions 
caused by the free-fall collapse over the time $t_s$:

\begin{equation} 
l(R,\theta)=\Omega_s (R \sin\theta)^2
\left(1+\frac{t_s}{t_{f\!f}(R)}\right)^{4/3},
\end{equation}

\begin{equation} B^r=\frac{B_0\sin \theta \cos \theta}{\sqrt{\gamma}}
R^2\left(1+t/t_{f\!f}(R)\right)^{4/3},
\end{equation}

\begin{equation} B^{\theta}=\frac{B_0\sin^2 \theta }{2\sqrt{\gamma}}
2R\left(1+t/t_{f\!f}(R)\right)^{1/3},
\end{equation} 
where $t_{f\!f}(R)=\sqrt{2R^{3}/9GM_h}$ is the local free fall
time scale and $\gamma$ is the determinant of the metric tensor of space (see
Appendix \ref{append}).  In these simulations we studied three different
cases summarised in Table \ref{table1}.

\begin{table}
\caption{Numerical models.}
\begin{tabular}{|c||c c c c c c|} \hline Model & $a$ & $B_{0}$ & $\Psi_{28}$ &
$L_{BZ}$ & $\dot{M}_h$ & $L_{BZ}/\dot{M}_{h}c^2$\\ \hline M1 & 0.6 & 1.4 &
0.46 & --- & --- & --- \\ M2 & 0.6 & 4.2 & 1.5 & 0.44 & 0.017& 0.0144 \\ M3 &
0.45 & 8.4 & 3.1 & 1.1 & 0.012& 0.049\\ \hline %\tabletext{}
\end{tabular}

$a$ is the black hole spin; $B_{0}$ is the initial
magnetic field strength in the units of $10^7$G; $\Psi_{28}$ is the magnetic
flux accumulated by the black hole by the time of explosion in the units of
$10^{28}\mbox{G}\,\mbox{cm}^2$; $L_{BZ}$ is the total power of the
BZ-mechanism during the explosion in the units of $10^{51}$ erg/s; 
$\dot{M}_{h}$ is the black hole mass accretion rate during the explosion in 
the units of $M_{\odot}$ s$^{-1}$.

\label{table1}
\end{table}

The simulations were carried out with 2D axisymmetric GRMHD code described in 
\citet{K99,K04}.  The gravity effects are introduced via the Kerr metric with fixed
parameters; the Kerr-Schild coordinates are used in order to avoid the
coordinate singularity at the horizon.  The computational grid is uniform in
the polar angle, $\theta$, where it has 180 cells and logarithmic in the spherical
radius, $R$, where it has 445 cells.  The inner boundary is located just
inside the event horizon and adopts the free-flow boundary conditions. The
outer boundary is located at $R=8.3\times 10^9 $cm and at this boundary the
flow is prescribed according to the Bethe's model.

In the simulations we used realistic equation of state (EOS) that takes into
account the contributions from radiation, lepton gas (including pair plasma),
and non-degenerate nuclei (hydrogen, helium, and oxygen).  This is achieved
via incorporation of the EOS code HELM \citep{TS00}, The neutrino cooling is
computed assuming optically thin regime and takes into account URCA-processes
\citep{IIN69}, pair annihilation, photo-production, and plasma emission
\citep{S87}, as well as synchrotron neutrino emission \citep{B97}. In fact,
URCA-processes strongly dominate over other mechanisms in this problem.
Photo-disintegration of nuclei is included via modification of the EOS following
the prescription given in \citet{ABM05}.  The equation for mass fraction of
free nucleons is adopted from \citet{WB92}. We have not included the radiative
heating due to annihilation of neutrinos and antineutrinos produced in the
accretion disk mainly because this requires elaborate and time consuming
calculations of neutrino transport.

%ssssssssssssssssssssssssssssssssss
\subsection{Results} 
%ssssssssssssssssssssssssssssssssss

\begin{figure}
\includegraphics[width=84mm,angle=-0]{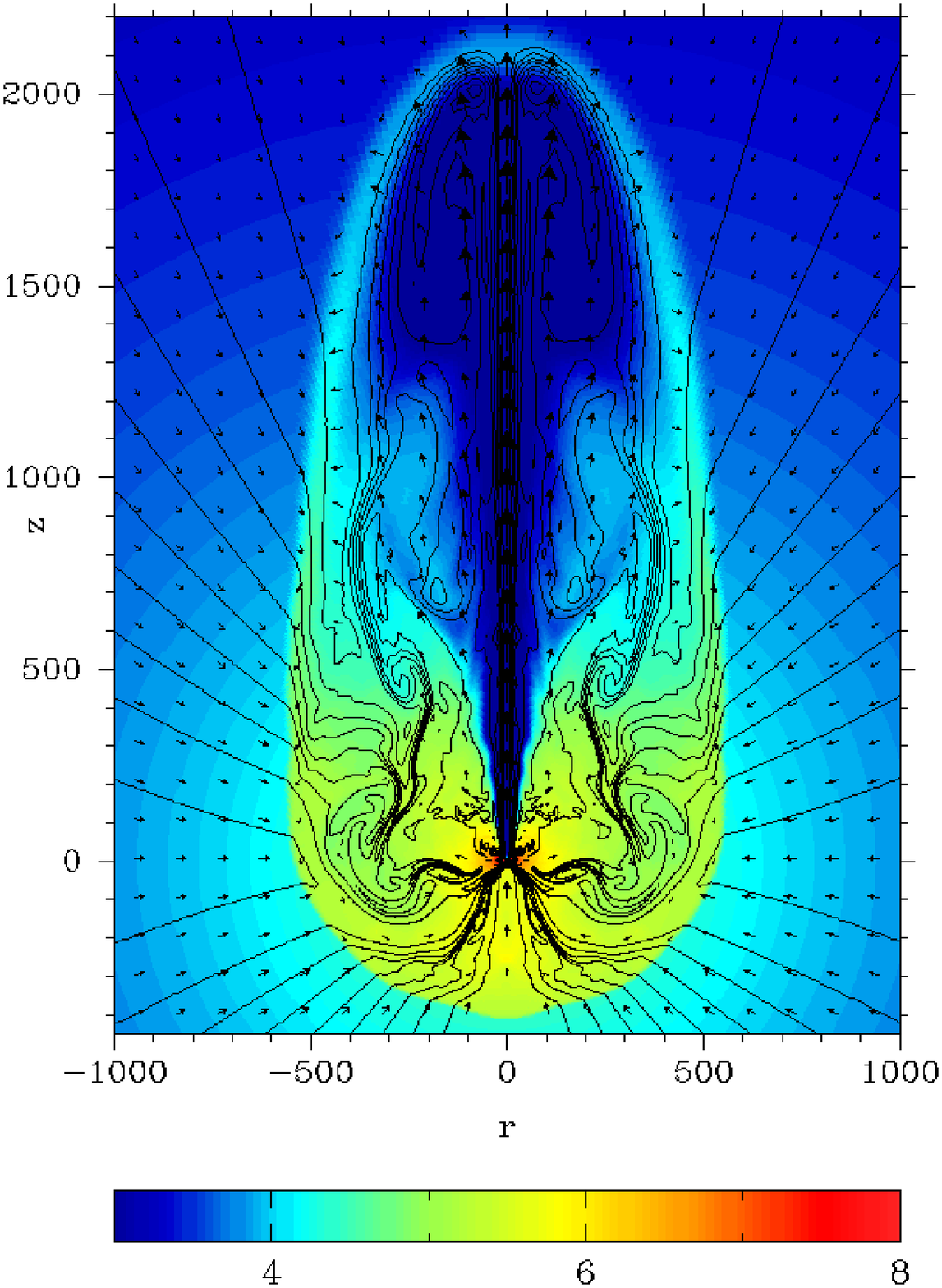}
\caption{ Model M2 ($B_0=2.2\times10^7$G, $a=0.6$) at the time of  $t=1.35$s 
after the start of simulations (18.3s after the start of the stellar collapse).
The colour image shows the baryonic rest mass density, $log_{10}\rho$, in CGS
units, the contours show the magnetic field lines, and the arrows show the
velocity field.}
\label{M2}
\end{figure}

\begin{figure}
\includegraphics[width=84mm]{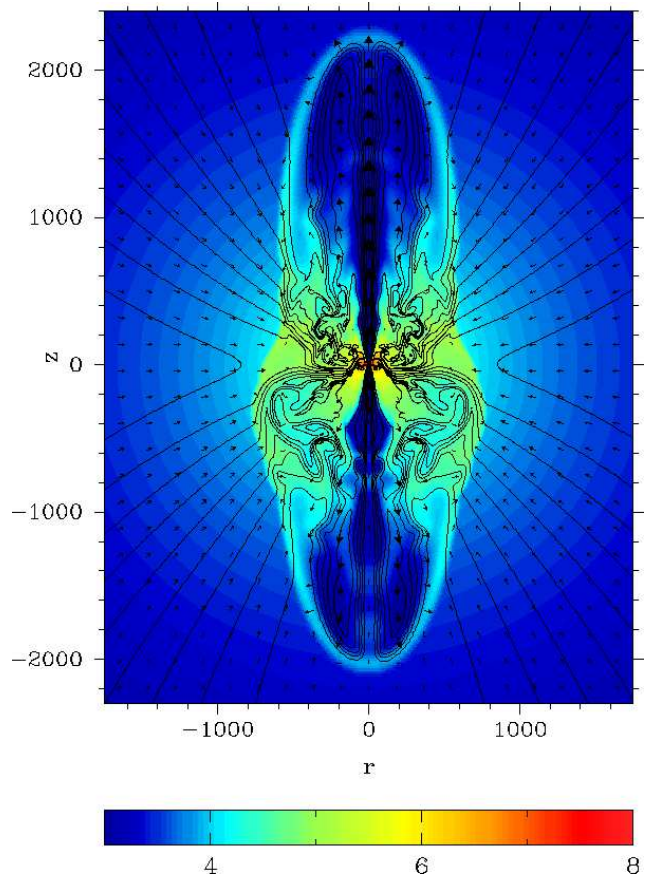}
\caption{ Model M3 ($B_0=8.8\times 10^{7}$G, $a=0.45$) at $t=1.35$s after the
start of simulations (18.3s after the start of the collapse). The colour
image shows the baryonic rest mass density, $log_{10}\rho$, in CGS units, the
contours show the magnetic field lines, and the arrows show the velocity
field.}
\label{M3}
\end{figure}

In general, the results of these simulations are in agreement with our
previous studies \citep{BK08,BK08b,KB09}. Because of the modified setup, which
corresponds to the later stages of the collapse, the accretion disk if formed
straight away. At the same time, the accretion shock separates from the disk
surface and quickly expands up to $R\simeq 100-200 R_g$. In the model M1, the
shock then begins to oscillate and no jets emerge by the end
of the simulations, $t=19.5$s. In contrast, both the models M2 and M3
eventually develop polar jets of relativistic plasma which are powered via the
Blandford-Znajek mechanism (see Figs.\ref{M2} and \ref{M3}).  These results
comply with the BZ activation condition, $\kappa\ge 0.2$, where

\begin{equation} 
\kappa = \frac{\Psi}{4\pi r_g \sqrt{\dot{M}c}},
\label{act}
\end{equation} 
is the activation parameter, $\Psi$ is the magnetic flux
threading the black hole and $\dot{M}$ is the mass accretion rate of the
collapsing star \citep{KB09}.  For the parameters of the present simulations we
have $\kappa\simeq 0.07 \Psi_{28}$, where the magnetic flux is given in the units
of $10^{28}\mbox{G}\,\mbox{cm}^2$, and, thus, one would expect the BZ mechanism to 
become activated for $\Psi_{28}>3$. As one can see from the data presented in 
Table 1 which was indeed the case.

According to the simple monopole model of black hole magnetosphere, the power
of the BZ mechanism is
\begin{equation} \dot{E}\sub{BZ}=1.4\times10^{51} f_2(a) \Psi\sub{28}^2
\left(\frac{M_b}{10 M_\odot}\right)^{-2} \,\mbox{erg}\, \mbox{s}^{-1},
\label{e-bz}
\end{equation}
where $f_2(a)=a^2\left(1+\sqrt{1-a^2}\right)^{-2}$ \citep{BK08}.  Like in our
previous simulations, the direct measurements of energy flux across the BH
horizon roughly agree with this result (see Table 1).

One significant difference with the results of previous simulations is the
development of a one-sided jet in model M2. Although noticeable deviations
from the equatorial symmetry have been observed before, in particular the
asymmetric oscillations of the accretion shock, such a strong deviation is
observed for the first time.  The initial solution is not exactly symmetric
because of the rounding errors, but they are tiny and the observed braking of
the equatorial symmetry has to be rooted in the nonlinear dynamics of the
flow.  It appears that the accretion flow, which is deflected towards the
equatorial plane at the oblique shock driven by the northern jet, protrudes
into the southern hemisphere. There it collides with the accretion flow of the
southern hemisphere and together they stream towards the black hole's southern
pole, thus suppressing the development of a southern jet.

If persistent, such a one-side jet could impart a strong kick on the black
hole and the binary, significantly altering its motion in the parent galaxy
\citep{F09}. The maximum kick velocity can be estimated as
\begin{equation} v_{kick} = \frac{E_{jet}}{cM_h} \approx 170
\left(\frac{E_{jet}}{10^{52}\mbox{erg}}\right) \left(\frac{10
M_{\odot}}{M_h}\right) \mbox{km}\,\mbox{s}^{-1},
\label{kick}
\end{equation} which is consistent with the observations of the X-ray binary
XTEJ1118+480 \citep{G05}.  However, at present we cannot say whether such
one-sidedness can persist during the life-time of LGRB or this is just a
transient phenomenon.

%fffffffffffffffffffffffffffffffffffffffffffffffffffffffffffffffff
%fffffffffffffffffffffffffffffffffffffffffffffffffffffffffffffffff

%fffffffffffffffffffffffffffffffffffffffffffffffffffffffffffffffff
%fffffffffffffffffffffffffffffffffffffffffffffffffffffffffffffffff

%%%%%%%%%%%%%%%%%%%%%%%%%%%%%%%%%%%%%%%%%%%%%%%%%%
\section{Merger Scenario}
%%%%%%%%%%%%%%%%%%%%%%%%%%%%%%%%%%%%%%%%%%%%%%%%%%
\label{CE}

The case of close tidally-locked binary considered above involves binaries
with orbital separation very close to the size of the Roche lobe of the WR
star and this suggests to go one step further and consider the case of even
smaller separation which can lead to the common envelope evolution resulting
in a merger of the binary \citep{ty79} and GRB explosion \citep{zf01}. Such a
merger can be divided into tree phases.  During the fist phase, the compact
companion spirals inside the extended envelope of the normal star and spins it up
via deposition of its orbital angular momentum. The compact star also
increases its mass and spin via the Bondi-type accretion. According to the
simulations of \citet{zf01}, during the last 500s of the in-spiral the compact
star can accumulate up to 3.5$M_\odot$. Thus, the mean accretion rate is less
than $10^{-2}M_\odot/$s implying inefficient neutrino heating.

The second stage begins when the compact star approaches the centre of its WR
companion and the accretion rate increases. \citet{zf01} find that in the case
of 16$M_\odot$ companion the neutrino annihilation mechanism can operate for
around 60 seconds and release about $10^{52}$erg.  This is more than enough to
drive a supernova explosion.  For the companion mass below $8M_\odot$ the
neutrino heating is too weak and the second phase is absent.

The third phase takes place if the second phase does not result in the 
supernova explosion or if the explosion is highly non-spherical and does not 
remove the equatorial layers of the WR star. 
During this phase the compact object, already a black hole,
accretes these layers, which have been spun up during the first
phase. Assuming that the mass of the compact star is small compared to the
mass of its WR companion, its orbital angular momentum can be found via the
Keplerian law
$$
   J_c(R)=M_c \sqrt{GM(R)R},
$$
where $M(R)$ is the WR mass inside the radius $R$. As the compact star moves
from the radius $R$ to $R+dR$, it transfers the angular momentum 
$dJ_c(R) =(dJ_c/dR)dR$ to the envelope of the WR star. Assuming that most of 
this angular momentum is transferred to the mass $dM=(dM/dR) dR$ of the envelope 
located between $R$ and $R+dR$, we obtain the 
specific angular momentum of the envelope after the merger as  
$$
l\simeq\frac{dJ_c}{dM}=\frac{dJ_c/dR}{dM/dR}.
$$ 
For the Bethe's model, where $M(R)$ is given by Eq.(\ref{M}), this gives 

\begin{equation} 
l \simeq \frac{M_c}{2} \left( \frac{GR}{M(R)} \right)^{1/2}
\left(1+\ln(R/R_c)\right), 
\label{lhe}
\end{equation} 
which is smaller than the local Keplerian angular momentum provided 
$M(R)>M_c(1+\ln{R/R_c})/2$. This suggests that if $M_s\gg M_c$ then only 
a small fraction of the common envelop is lost during the merger. For
$R= R_s$ this equation gives

\begin{equation} 
l \simeq 5.2\times 10^{18} \bfrac{M_c}{2M_\odot}
\bfrac{R_s}{R_\odot}^{1/2} \bfrac{M_s}{10M_\odot}^{-1/2}\!\!\!\!
\mbox{cm}^2\mbox{s}^{-1}.
\label{lhe1}
\end{equation} 
In the $\alpha$-model, the accretion time scale of the disk
with such angular momentum can be estimated via
%\begin{equation}
$$
t_d \simeq \frac{1}{\alpha \delta^2} \frac{l^3}{(GM_s)^2} =
$$
%\end{equation}
\begin{equation} \quad \simeq 8000\, \mbox{s} \bfrac{\alpha
\delta^2}{0.01}^{\!-1}\!\!\!  \bfrac{R_s}{R_\odot}^{3/2} \!\!\!
\bfrac{M_c}{2M_\odot}^{2} \!\!\!  \bfrac{M_s}{10M_\odot}^{-7/2}\!\!\! .
\label{t_d}
\end{equation} This is significantly longer than the duration of the stellar
collapse (see Eq.\ref{te}). In fact, such a
long time scale suggests the possibility of explaining the phase of shallow
decay and late flares in the X-ray light curves of LGRBs discovered by {\it
Swift} \citep{z07,ch07}.

To find the mass accretion rate as a function of time we note that
$$
\dot{M} = \frac{dM}{dt_d}=\frac{dM/dR}{dt_d/dR}.
$$ 
Using Eqs.(\ref{M},\ref{t_d},\ref{lhe}) to evaluate $dM/dR$ and $dt_d/dR$ we
obtain\footnote{This equation is the same as Eq.\ref{M-dot} but $t$ spans a
different range of timescales, now dictated by the disc accretion time.}

\begin{equation} 
\dot{M} \simeq \frac{2}{3} \frac{M_s}{\ln(R_s/R_c)}
\frac{1}{t} \simeq 1.45 \bfrac{M_s}{10M_\odot} \bfrac{t}{1\mbox{s}}^{-1}
\frac{M_\odot}{\mbox{s}}.
\label{M_dot}
\end{equation} 
Thus, on the time scale of $10^3-10^4$s the mass accretion rate 
is very low, $\dot{M}\simeq 10^{-3}\div 10^{-4} M_\odot\mbox{s}^{-1}$, 
ruling out the neutrino mechanism and leaving the BZ
mechanism clear favourite.  Indeed, the maximum possible amount of
magnetic flux that can be accumulated by the black hole is given by the
balance of magnetic pressure and the gas pressure of the accretion disk,

\begin{equation} 
\frac{B^2_{max}}{8\pi} \simeq P_g \simeq \rho c_a^2,
\label{mf-est}
\end{equation} 
where $c_a$ is the sound speed. If we utilize the model of
$\alpha$-disk and estimate the magnetic field strength at the gravitational
radius then the corresponding magnetic flux will be

\begin{equation} \Psi_{max} \simeq 3\times10^{29}
\bfrac{\alpha\,\delta}{0.03}^{-1/2} \!  \bfrac{M_b}{10M_\odot} \dot{M}_1^{1/2}
\mbox{G}\,\mbox{cm}^2,
\label{flux_max}
\end{equation} 
where $\dot{M}_1$ is the mass accretion rate in the units of
$M_\odot/$s.  Even for $\dot{M}_1$ as small as $10^{-4}$ this equation gives
the substantial value of $\Psi_{max}\simeq 3\times10^{27}
\mbox{G}\,\mbox{cm}^2$. The corresponding BZ power,
$\dot{E}_{BZ}\simeq 2.2\times 10^{49} \mbox{erg}/\mbox{s}$, is more than
sufficient to explain the X-ray observations, allowing the magnetic field to
be even weaker compared to the value suggested by Eq.\ref{mf-est}.

%%%%%%%%%%%%%%%%%%%%%%%%%%%%%%%%%%%%%%%%%%%%%%%%%%
\section{Discussion and Conclusions}
%%%%%%%%%%%%%%%%%%%%%%%%%%%%%%%%%%%%%%%%%%%%%%%%%%
\label{DC}

One of the main issues of this study was to investigate the efficiency of the
tidal spin up in close massive binaries in the context of the collapsar model
of LGRBs. In particular, we wanted to find out the typical masses and spins of
the black holes formed during the collapse of WR companion.  
It turns out, that the BH spin in this model is rather modest.  For example, 
in the most optimistic case of a binary with the smallest 
possible orbital separation, the spin parameter of the WR star is relatively high, 
$a_s\simeq 6$, and one may have expected the BH to be rapidly rotating.   
However, we find that the spin parameter is only $a \simeq 0.4$
at the time of the accretion disk formation, and $a\simeq0.8$ by the end of the
stellar collapse, which is significantly lower than the maximally possible
value $a=1$. This is mainly due to the
significant loss of angular momentum suffered by the mass accreted via the disc; 
the rejected angular momentum is either stored in the remote
part of the accretion disk or removed by a disk wind. Indeed, as
soon as the accretion disc is formed, the rate of accretion of angular
momentum slows down significantly.  Moreover, by the time of the disk
formation the black hole mass is already rather high, exceeding half of the
progenitor mass prior to the collapse.  Thus, the black hole simply runs out
of accreting matter before its rotation can approach the maximal possible rate
\citep[cf.][]{th74}.  

%Soon after the disk is formed the total mass accretion rate is dominated by 
%the disk and the accretion through the polar columns is not important.  
%This is why the third factor you have suggested is less important.

The mass accretion rate in this scenario is much lower compared to the 
usual $\dot{M} = (0.1\div1) M_{\odot}\,\mbox{s}^{-1}$ invoked in the 
standard collapsar model \citep{mw99}. This makes the
neutrino mechanism less attractive compared to the magnetic mechanisms, 
and the BZ-mechanism in particular. In fact, the very rapid decline
in the efficiency of neutrino mechanism below $\dot{M} \le 0.02-0.05
M_{\odot}\,\mbox{s}^{-1}$\citep{pwf99,ZB09} makes the explanation of the LGRB
bursts with duration $\ge 100$s rather problematic even within the standard
collapsar model due to the low mass accretion rate expected on such time scale
(see Eq.\ref{M_dot} and Fig.\ref{mod_AB_mdot}).

However, the BZ mechanism could have its own difficulties in this scenario.
Indeed, it requires very strong ordered magnetic field.  For example, in order 
to provide the power of $10^{50}\mbox{erg}\,\mbox{s}^{-1}$ the black hole of 
mass $10M_{\odot}$ and $a=0.6$ should accumulate the magnetic flux of order
$\Psi=8\times10^{27}\mbox{G}\,\mbox{cm}^2$.  The magnetic flux necessarily to
activate the BZ mechanism soon after the formation of accretion disk is 
even higher.  According to Table I this is of the
order few$\times 10^{28}\mbox{G}\,\mbox{cm}^2$.  Perhaps the free-fall
accretion rate set up in our simulation is a bit too high and could have been
reduced by a factor of 10. However according to Eq.(\ref{act}), this would
reduce the critical value of magnetic flux only by a factor of 3.

The origin of such strong field is not clear. It could be generated via 
magnetic dynamo in the accretion disc \citep[e.g.][]{bran95} or in the
convective core of the progenitor \citep[e.g.][]{cm01}. It may also be 
inherited by the progenitor from the interstellar medium (ISM) during its 
formation \citep[e.g.][]{BS04}.  The current status of both the stellar and
disk dynamo theories does not really allow to make reliable
conclusions. Even the issue of advection of externally generated magnetic 
field by the accretion disk onto the central black hole is still unresolved
\citep[e.g.][]{bkr74,van89,gg09,su05,lov08,i08}. There seems to be a general 
agreement that accretion disks produce mainly azimuthal magnetic 
field and unable to generate poloidal field on scales exceeding 
the disk hight. The magnetic dynamo in convective cores of massive
stars could be more promising in this respect. For example, from the 
results of \cite{cm01} it seems possible to generate up to 
$\Phi\simeq 10^{28}\mbox{G}\,\mbox{cm}^2$ in the convective cores of B stars.

By design, our numerical simulations cannot address the issue of magnetic
field generation is accretion disks and strictly-speaking deal only with the
fossil model of magnetic field. The numerical results by \citet{BS04} suggest
that strong fossil field can relax to a simple ordered configuration with
dipolar poloidal field on a relatively short time scale, which makes our setup
not that unrealistic. However, further studies are required to verify this
model.  The observations of massive stars do not support magnetic flux of 
order $10^{28}\mbox{G}\,\mbox{cm}^2$ and higher.  The current record is held 
by $\theta^1$ Ori C, whose dipolar magnetic flux
$\Psi\simeq 2\times10^{27}\mbox{G}\,\mbox{cm}^2$, \citet{don02}).  One may
speculate that most of the magnetic flux is hidden in the stellar interior.  
Indeed, the resistive time scale across the extended radiative outer layers of
massive stars exceeds their life time by many orders of magnitude \citet{BS04}.  

The fact that the magnetic flux of neutron stars is less than 
$10^{27}\mbox{G}\,\mbox{cm}^2$ also seems to be working against the fossil
hypothesis.  However, neutron stars are collapsed compact Fe cores of massive 
stars. The typical cross section of such a core is several orders of magnitude 
below that of the whole star and, thus, the core may account only for a 
small fraction of the total magnetic flux hidden inside the supernova progenitor.

The host galaxies of LGRBs show strong evidence of enhanced star formation
\citep{BDKF98,S01,Fr06}. It is interesting that the recent observations of
such starburst galaxies also indicate strong ISM magnetic field, in fact up to
ten times stronger compared to the Milky Way \citep{bk05,b08}. This suggests
that magnetization of young stars in the host galaxies of LGRBs can be 
abnormally high as well. 

Another interesting proposal stems from the theory of Sun's magnetic activity
proposed by \citet{uz07}. In particularly, he argued that fast reconnection
can only operate in collisionless plasma and in the collisional regime the
reconnection rate reduces to the much slower rate of Sweet-Parker. 
Since the collapsar plasma is collisional even in the rarefied funnel of 
the accretion disk then, according to this theory, the reconnection rate 
in the black hole magnetosphere can be relatively slow. 
An additional unexplored factor in the LGRB context is the effects of quantum
physics on magnetic reconnection. Indeed, the expected magnetic field strength
is well above the quantum value of $B_q=m_e^2 c^3/\hbar e = 4 \times
10^{13}$G.  One may speculate that under this conditions the reconnection rate
becomes even slower. 

In the case of slow reconnection, the black hole may be 
able to build strong magnetic field via collecting the alternating magnetic field
generated in the accretion disk. Since the magnetic stresses are invariant
with respect to change of magnetic polarity such striped structure of magnetic
field has no effect on the efficiency of the BZ-mechanism.  Further downstream
of the LGRB flow, where its plasma becomes collisionless or the magnetic field
becomes sufficiently weak, the reconnection accelerates.  However, as long as
this occurs beyond the Alfven surface, which for a black hole with reasonable
spin does not greatly exceed the gravitational radius, this does not disrupt
the near magnetosphere of the black hole and does not reduce the efficiency
of the BZ-mechanism. Moreover, such delayed reconnection could promote bulk
acceleration of the LGRB flow \citep{DS02}.

Finally, the neutrino heating of the polar region, not included in our
analysis and simulations, may also play a very important role, by initiating
the LGRB outflow and creating the low density chanel in the polar direction
early on, when the mass accretion rate is still sufficiently high for
effective neutrino-antineutrino annihilation. This would allow the
BZ-mechanism to be activated along the field lines filling the channel even if
the black hole magnetic flux is much lower compared to the values quoted
above.  Later on, when the mass accretion rate drops and the neutrino
mechanism can no longer provide sufficient power, the BZ-mechanism can take
over the role of main driver of the LGRB flow. One may even contemplate the
scenario where the GRB precursors are related to the neutrino-driven stellar
explosions and the main bursts to the magnetically-driven BH jets unleashed in
the space cleared up by the blast. The delay between the two phases could be
related to the disruption of the accretion flow by the supernova blast
\citep{WM07}. Because of the rotational effects the disruption may not be as
severe in the equatorial direction, compared to the polar direction, as in 
the one-dimensional simulations
by \citet{MWH01}, leading to shorter fallback time-scales. The magnetic jets,
though very powerful, could be less disruptive compared to the neutrino-driven
jets because the magnetic hoop stress, associated with the azimuthal component 
of magnetic field, makes the sideways expansion of the jet cocoon less 
effective.

The most interesting, in view of the recent {\it Swift} observations of LGRB
afterglows, version of the close binary scenario for GRB progenitors is the
common envelope case, where the compact star, either a black hole from
the onset or a neutron star which eventually collapses into a black
hole, spirals inside the normal WR star. The large angular momentum
transferred to the external layers of the WR star quite naturally leads to
long accretion time scales, $\simeq 10^4$s.  Thus, the central engine of LGRB
jets arising in this scenario could operate for a sufficiently long time to
explains the shallow phase of the X-ray light curves discovered by {\it Swift}
\citep{z07}. The X-ray flares, which are often seen during this phase, may
result from the gravitational instabilities developing in this disc
\citep{paz06}.  Although the BZ mechanism is not that sensitive to the mass
accretion rate as the neutrino mechanism, some dependence is still expected.
For example, Eqs.(\ref{e-bz},\ref{flux_max}) suggest that the power of the BZ
mechanism may be proportional to the mass accretion rate.  This can explain
why the gamma ray emission becomes undetectable on the time-scale of the
shallow decay of X-ray afterglows.

The extremely high rotation rates, about $50\%$ of the break-up speed, assumed
in the single progenitor model by \citet{YL05} and \citet{WH06} imply that in
this model the outer layers of the collapsing star can also develop
long-lived accretion disk. Indeed, in the ``showcase'' model 16TI of
\citet{WH06} the outer $\simeq 2 M_\odot$ have the specific angular momentum
increasing outwards from $10^{18}$ to $10^{19}$cm$^2$s$^{-1}$ at the
pre-supernova phase.  According to Eq.\ref{lhe1} this corresponds to the disc
accretion time scales of order $10^4$s.  However, such a long time scale still
rules out the neutrino annihilation as the mechanism for powering the collapsar 
jets.

%%%%%%%%%%%%%%%%%%%%%%%%%%%%%%%%%%%%%%%%%%%%%%%%%%%%%%%%%%%%%%%%%
\section*{Acknowledgments}
%%%%%%%%%%%%%%%%%%%%%%%%%%%%%%%%%%%%%%%%%%%%%%%%%%%%%%%%%%%%%%%%%
This research was funded by STFC under the rolling grant ``Theoretical
Astrophysics in Leeds'' (SSK and MVB). The numerical simulations were carried
out on the St Andrews UK MHD cluster and on the Everest cluster of the White
Rose Grid.  Both authors are very grateful to the Nordic Institute for
Theoretical Physics (NORDITA), where this work was completed, as well as to
the organisers of the NORDITA research program ``Physics of Relativistic
Flows'' for hospitality and generous support.

\appendix

\section{Evolution of angular momentum and magnetic field in the Bethe's model 
of stellar collapse}
\label{append}

The free fall model by \citet{b90} approximates the kinematics of stellar
collapse by the model
$$
\frac{dR}{dt} = \left\{
\begin{array}{rcl} 0 & \mbox{if} & t\le 0,\\ -(2GM(R)/R)^{-1/2} & \mbox{if} &
t>0,
\end{array} \right.
$$
where $R$ is the radius of collapsing shell and $M(R)$ is the mass inside this
radius. Since $dM/dt=0$, this equation is easily integrated
$$
   R_0(R,t) = R (1+t/t_{f\!f}(R))^{2/3} = \mbox{const},
$$
where $R_0=R(0)$ and $t_{f\!f}(R)=\sqrt{2R^3/9GM(R)}$ is the local free fall
time.  

Given the initial distribution of angular momentum, $l_0=\Omega
(R_0\sin\theta)^2$, the conservation of angular momentum yields
$$
 l(R,t)=l_0(R_0(R,T)) = \Omega_s (R\sin\theta)^2 (1+t/t_{f\!f}(R))^{4/3}.
$$
Similarly, the conservation of magnetic flux requires
$$
 \Psi(R,\theta,t) = \Psi_0(R_0(R,t),\theta).
$$
For the uniform initial magnetic field,
$$
 \Psi_0(R_0,\theta) = \pi B_0 \sin^2\theta R_0^2.
$$ 
Thus,
$$
   \Psi(R,\theta,t) = \pi B_0 \sin^2\theta R^2 (1+t/t_{f\!f}(R))^{4/3}.
$$
The poloidal magnetic field can be found via
$$
B^i_{p}=\frac{1}{2\pi}e^{ij\varphi}\partial_j\Psi,
$$
where $e^{ijk}$ is the Levi-Civita tensor of space. This gives us

\begin{equation} B^r=\frac{B_0\sin\theta \cos\theta}{\sqrt{\gamma}}
R^2\left(1+\frac{t}{t_{f\!f}(R)}\right)^{4/3}
\label{br}
\end{equation} and

\begin{equation} B^{\theta}=\frac{B_0\sin^2\theta }{\sqrt{\gamma}}
R\left(1+\frac{t}{t_{f\!f}(R)}\right)^{1/3},
\label{bt}
\end{equation} where $\gamma$ is the determinant of the metric tensor of space
and the vector components are given in the non-normalised coordinate basis,
${\partial/\partial x^i}$. This approach has been used in \citet{bkr74}.
\end{document}